%% file: main.tex
\documentclass[aps,floatfix,prd,nofootinbib,superscriptaddress,reprint,showpacs,10pt,preprintnumbers,longbibliography]{revtex4-1}
\usepackage[T1]{fontenc}
\usepackage[utf8]{inputenc}
\usepackage[pdftex]{graphicx}
\usepackage[UKenglish]{babel}
\usepackage{float}
\usepackage{amsmath}
\usepackage{amssymb}
\usepackage{mathtools}
\usepackage{siunitx}
\usepackage{amsfonts}
\usepackage{dsfont}
\usepackage{array}
\usepackage{bm}
\usepackage{mathrsfs}
\usepackage{pifont}
\usepackage{multirow}
\usepackage{upgreek}
\usepackage[dvipsnames]{xcolor}
\usepackage[lined,boxed,commentsnumbered]{algorithm2e}
\usepackage{nameref,zref-xr}
\usepackage{csquotes}
\usepackage[pdftex,
  pdftitle={Beer Mats make bad Frisbees},
  pdfauthor={Johann Ostmeyer, Christoph Schürmann, Carsten Urbach},
  bookmarks,
  colorlinks,
  linkcolor=myblue,
  citecolor=mymagenta,
  menucolor=black,
  urlcolor=myblue,
  plainpages=false,
  pdfpagelabels,
  hypertexnames=false]{hyperref}
\usepackage{cleveref}

\DeclareMathOperator{\atan}{atan2}
\newcommand{\del}[2]{\ensuremath{\frac{\partial #1}{\partial#2}}}
\newcommand{\matr}[2]{\left(\!\begin{array}{#1}#2\end{array}\!\right)}
\newcommand{\eto}[1]{\ensuremath{\mathrm{e}^{#1}}}
\newcommand{\md}{\ensuremath{\mathrm{d}}}

\newcommand{\defined}[3]{\ensuremath{#1\in#2 &: \text{#3}}}
\newcommand{\constant}[3]{\ensuremath{#1=#2 &: \text{#3}}}

\newcommand{\st}{\ensuremath{\sin\theta}}
\newcommand{\ct}{\ensuremath{\cos\theta}}
\renewcommand{\sp}{\ensuremath{\sin\phi}}
\newcommand{\cp}{\ensuremath{\cos\phi}}
\newcommand{\dphi}{\ensuremath{\dot{\phi}}}
\newcommand{\dtheta}{\ensuremath{\dot{\theta}}}
\newcommand{\dpsi}{\ensuremath{\dot{\psi}}}
\newcommand{\pphi}{\ensuremath{p_{\phi}}}
\newcommand{\ptheta}{\ensuremath{p_{\theta}}}
\newcommand{\ppsi}{\ensuremath{p_{\psi}}}

\newcommand{\vek}{\boldsymbol}
\newcommand{\vd}{\ensuremath{{\vek v}\cdot {\vek D}}}

\definecolor{mymagenta}{RGB}{200, 0, 100}
\definecolor{myblue}{RGB}{45, 48, 146}

\graphicspath{{figures/}}

\begin{document}
\title{Beer Mats make bad Frisbees}

\author{Johann Ostmeyer}
\affiliation{Helmholtz-Institut für Strahlen- und Kernphysik, University of Bonn, Nussallee 14-16, 53115 Bonn, Germany}
\affiliation{Bethe Center for Theoretical Physics, University of Bonn, Nussallee 12, 53115 Bonn, Germany}

\author{Christoph Schürmann}
\affiliation{Max-Planck-Institut für Radioastronomie, Auf dem Hügel 69, 53121 Bonn, Germany}
\affiliation{Argelander-Institut für Astronomie, Universität Bonn, Auf dem Hügel 71, 53121 Bonn, Germany}

\author{Carsten Urbach}
\affiliation{Helmholtz-Institut für Strahlen- und Kernphysik, University of Bonn, Nussallee 14-16, 53115 Bonn, Germany}
\affiliation{Bethe Center for Theoretical Physics, University of Bonn, Nussallee 12, 53115 Bonn, Germany}

\begin{abstract}
  In this article we show why flying and rotating beer mats, CDs, or
  other flat disks will eventually flip in the air and end up flying with
  backspin, thus, making them unusable as frisbees. The crucial effect
  responsible for the flipping is found to be the lift attacking
  not in the center of mass but slightly offset to the forward
  edge. This induces a torque leading to a precession towards 
  backspin orientation. An effective theory is
  developed providing an approximate solution for the disk's
  trajectory with a minimal set of parameters. Our theoretical results
  are confronted with experimental results obtained using a 
  beer mat shooting apparatus and a high speed camera. Very good
  agreement is found.
\end{abstract}

\maketitle

\allowdisplaybreaks[1]

\input{sections/intro}

\input{sections/qualitative_explanation}
\input{sections/effective_theory}

\input{sections/experiments}

\input{sections/results}

\input{sections/conclusion}

\begin{acknowledgments}
  This work has been inspired by a trip to Munich in March 2017 with
  the Physics Show Bonn. We thank all the participants for this
  inspiration and the very enjoyable time together. We would also like
  to thank Michael Kortmann who lent us the high speed camera, helped
  us with the experiments and without whom we three theoreticians
  would have been completely lost. We thank Karl Maier for very useful
  discussions on the shooting apparatus' design. Last but not least we
  owe huge thanks to the mechanics and electronics labs of the HISKP
  Bonn and in particular Dirk Lenz for building the beer mat shooting
  apparatus. 
  
  This work
  is supported in part by the Deutsche Forschungsgemeinschaft (DFG,
  German Research Foundation) and the 
  NSFC through the funds provided to the Sino-German
  Collaborative Research Center CRC 110 “Symmetries
  and the Emergence of Structure in QCD” (DFG Project-ID 196253076 - TRR 110, NSFC Grant No. 12070131001).
  
  Our sincere apologies to everyone hit by a beer mat, be it through inaccurate aim or due to us instigating others to perform silly experiments.
\end{acknowledgments}

\bibliography{bibliography}

\onecolumngrid
\appendix

\input{sections/forces}

\input{sections/numerical_solution}
\input{sections/more_verification}
\input{sections/experimental_setup}
\input{sections/no_spin}

\end{document}

%% file: sections/intro.tex
\section{Introduction}

\begin{figure}[t]
  \centering	
  \includegraphics[trim={1.7cm 1.2cm 1.7cm 1.2cm},clip, width=.9\columnwidth]{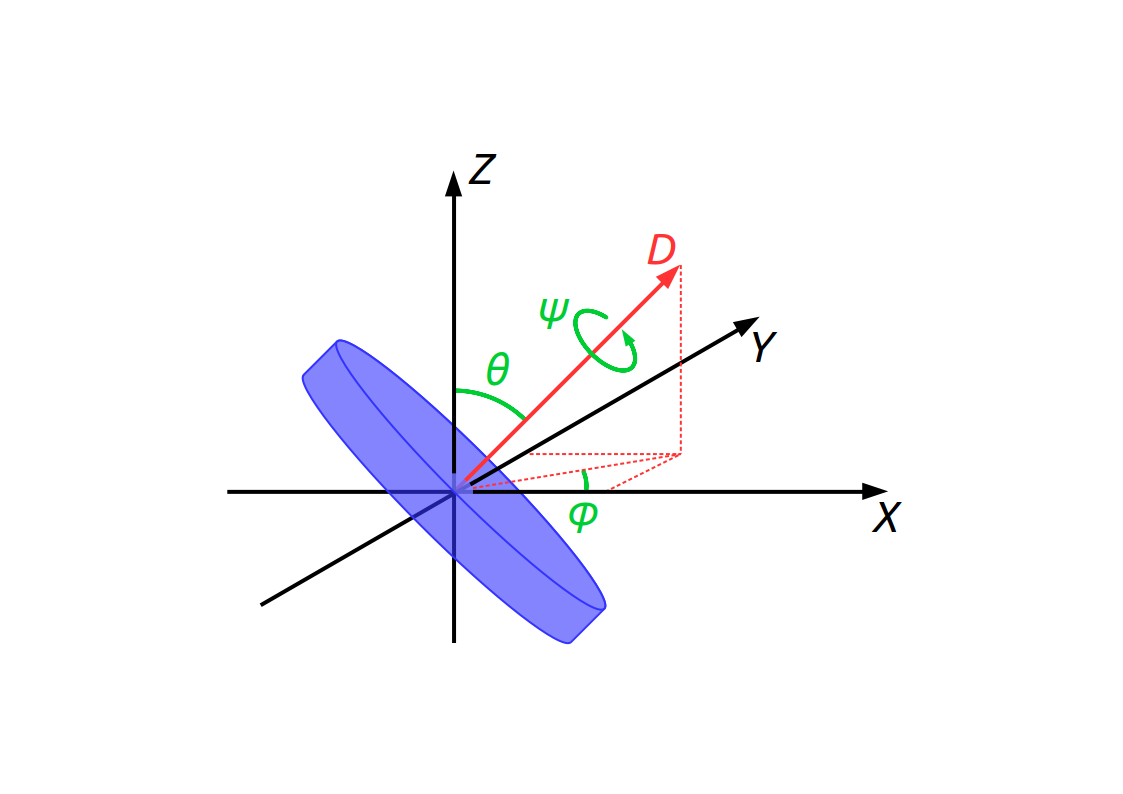}
  \caption{Sketch of the disk and the most relevant
    coordinates.
  }\label{fig_disk_sketch}
\end{figure}

A beer mat (also known as drink coaster) is a commodity most elegantly used to rest a glass on in
order to protect a table surface~\cite{Cambridge}. However, not only for a
physicist there are more exciting applications for a beer mat, one of
which is to let it fly. Usually, it is a circular piece of cardboard
with a diameter of about $\SI{10}{\centi\meter}$, though there are
also other shapes in use. For the following we will idealise it as a
disc with radius $r$, mass $m$ and negligible thickness, for a
sketch see \Cref{fig_disk_sketch}.

If one tries to throw a beer mat, one quickly realises that one can
only achieve reasonable flight distances if the mat rotates around the
axis perpendicular to it's circular area, depicted as $\vek D$ in
\Cref{fig_disk_sketch}. The such generated angular 
momentum stabilises the orientation of the disk via angular momentum
conservation preventing chaotic rotations around one of the two other
rotation axes of the disc, for which it is known from classical
mechanics that rotations around these are in practice unstable, given
the relatively small mass of the beer mat.

Now one could expect the mat to fly similarly to a frisbee, i.e.\ with
angular momentum pointing up or down, called `\textit{sidespin}' from
now on, however, it still turns out to 
be difficult to predict the path of the flying mat: a seemingly random
time $\tau$ after the flight started the mat begins to either turn
left or right, depending on it's direction of rotation, and -- if it
does not hit ground beforehand -- ends up flying with backspin (i.e.\ 
with rotation axis pointing sidewards perpendicular to
the direction of flight with the upper side rotating against the
direction of flight). With
a few more experiments one realises that starting the flight with
backspin  is stable, while a flight with topspin is not.

The observation of the aforementioned seemingly random times $\tau$
might lead one to hypothesise there to be a chaotic effect in the
flight of a beer mat. However, it will turn out that this effect comes
from the inability of humans to throw the disk reproducibly.

In principle, the theory of a rotating thin disk moving in air is
known. Unfortunately, solving the corresponding equations including effects
from turbulence is analytically intractable and even numerically
highly demanding. An effective treatment is, therefore, in order by
treating the disk as a point-like object to a large extend, which
allows one to avoid to solve the full fluid dynamics equations. The
closest to the situation we are considering here comes the literature
about the flight of a frisbee, see for
instance~\cite{frisbee_thesis,Hubbard00simulationof,honors_theses,Lorenz_2005,Motoyama2002ThePO}\footnote{The
literature on this topic is sparse and often not published in peer
reviewed journals.}.
However, a frisbee weighs significantly more and has modified edges to
stabilise it's flight characteristics such that the effects that we
will study here are sufficiently suppressed to be neglected. 

The aim of this paper is threefold: we will give a qualitative
explanation of the observed behaviour in the following
section~\ref{sec:qualitative}. From this section the understanding of
the phenomenon should be possible 
without much expert knowledge. Next, we will derive an
effective formalism based on a few assumptions describing the flight
of a beer mat. This formalism allows us to make predictions, which can
be tested experimentally. Thus, in the third part of this paper
we present experimental results. For the experiments to be
reproducible we have designed and constructed an apparatus which
allows us to throw beer mats with variable angular and forward
momentum. A high-speed camera is used to record the mats' flights and
the recordings are used to reconstruct the corresponding flight
trajectories. These trajectories are then compared to the predicted
trajectories from our effective theory.

Eventually, we compare the experimental results to our theory 
predictions, after a number of parameters are fitted to the
experimental data. The predictions are well confirmed, validating the
assumptions the effective theory was built on. This allows us to
generalise our findings to other types of disks, which could be tested
experimentally.

%% file: sections/qualitative_explanation.tex
\section{Qualitative Discussion}
\label{sec:qualitative}

\begin{figure*}[t]
  \centering
  \includegraphics[width=\textwidth]{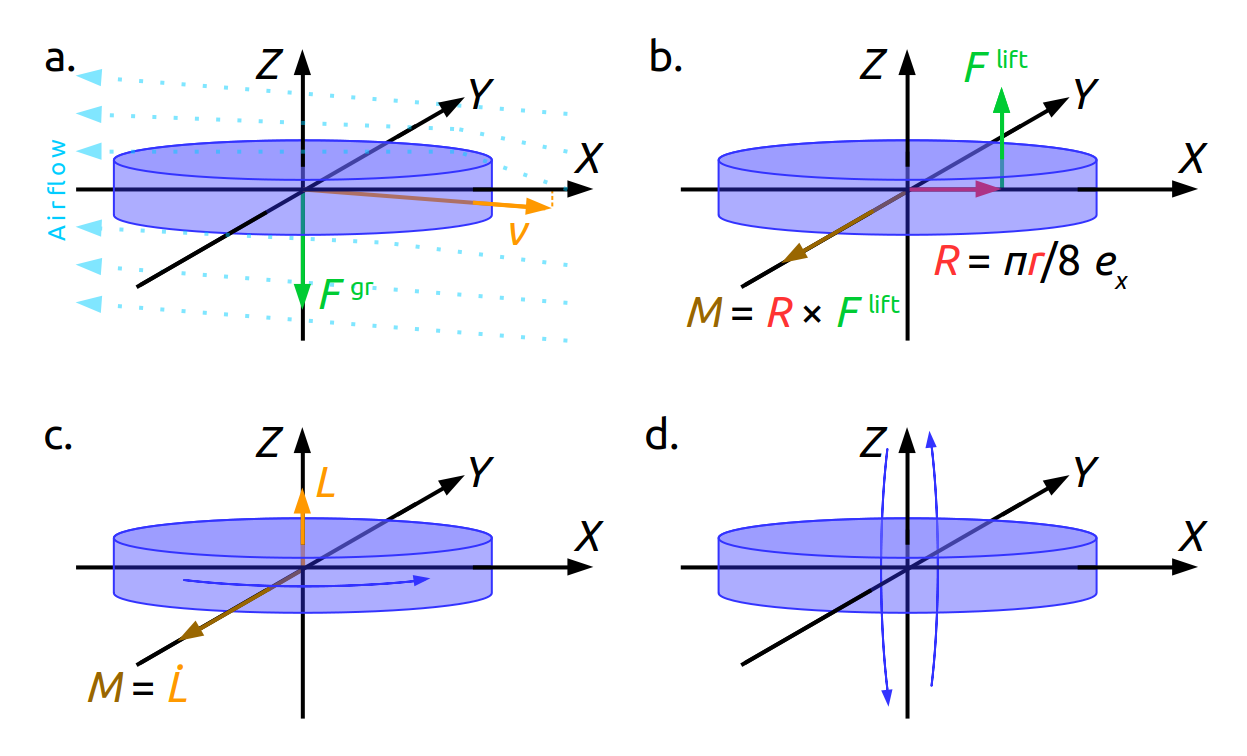}
  \caption{a. Airflow around the disk. b. Forces acting on the disk. c. Angular momentum and orientation of rotation. d. Torque forcing disk into backspin. }\label{fig_orientation_change}
\end{figure*}

Let us start in a situation where a spinning disk\footnote{The angular
  momentum is assumed to be large enough to guarantee a stable
  flight. See \cref{app:no_spin} for more
  information.}
with its angular momentum $\vek L$ perpendicular to the disk's surface
is moving through a medium (air) without gravitational force. Drag
will cause the disk to orientate itself such that it minimises 
air resistance by minimising its surface exposed to the airflow. This 
means the axis $\vek D$ is perpendicular to the flight direction.
Thus, any orientation with slightly tilted rotation
axis compared to the situation above will automatically adjust back.
In the equilibrium state no
torque is available to change the direction of angular momentum.

However, if one now includes the gravitational force pointing into the negative
$z$-direction, the disk will be accelerated downwards. Hence, the airflow approaches 
it slightly from below (see \Cref{fig_orientation_change}, A). This induces a lifting force 
$\vek{F}^\mathrm{lift}$ upwards~\cite{aerodynamik}. This force 
now does not attack the disk at its centre, but $R = \pi r/8$ towards the leading edge~\cite{abbott1959theory}.
Intuitively, this can be understood by the air stream being broken near the front edge of the disk, so that the lift is acting stronger at the front.
Therefore, $\vek{F}^\mathrm{lift}$ induces a torque $\vek{M} = \vek{R} \times \vek{F}^\mathrm{lift}$ which will cause a 
precession of the disk (\Cref{fig_orientation_change}, C).\footnote{The precession also causes the disk to change 
its flight direction because it accelerates towards the down-tilted edge. 
This effect, the so called `fade'~\cite{fade_explained}, is well known for frisbees.}
The lifting force and, thus, the torque vanishes only in back- or topspin position. But,
since the direction of this precession is such that the disk
approaches backspin orientation independently of the direction of
the angular momentum, the topspin orientation is meta stable and the
backspin orientation stable.

Therefore, if the flight of the disk is started not in backspin
position, one expects it to perform an oscillation around the backspin
position. This oscillation, however, is strongly damped in any real world experiment.

A secondary but nonetheless important stabilising effect is the Magnus force~\cite{magnus_effect}.
Together with drag it is responsible for the aforementioned damping.
The Magnus force acts orthogonal to both flight direction and rotation axis $\vek D$.
It does not introduce any torque and, thus, cannot be primarily
responsible for the trend towards backspin.
However, it can change the flight direction in such a way that backspin is preferred.
Imagine the spinning disk simply being dropped without initial spacial velocity
and $\vek D$ pointing in the $y$-direction, i.e.\ the leading edge pointing precisely
downwards. Once the disk begins to fall, it is accelerated
in the $x$-direction by the Magnus effect.%
\footnote{This has been visualised
	many times with a good example of backspin flight
	in~\cite{magnus_basketball}.}
The sign of this acceleration in $x$-direction prefers backspin again.
Small deviations from backspin can therefore be corrected by the Magnus effect adjusting the flight direction.

%% file: sections/effective_theory.tex
\section{Effective theory}
\label{sec:effective_theory}

In this section we will translate the qualitative arguments developed
in the previous section into formulae effectively describing the
flight of a beer mat (or any other thin disk). Thereafter, we will
test these formulae experimentally in the following section.

Note that this set of equations is not formulated with the goal to
explain the reality. Rather it is a means to predict movements with
reasonable precision. Terms without an immediate physical intuition
have to be read in this light. 

\subsection{Equations of motion}\label{sec:eom}

For a thin disk with radius $r$ and mass $m$ the element of the moment
of inertia tensor for rotations around the symmetry axis $\vek D$ (see
\Cref{fig_disk_sketch}) reads $I_{D} = mr^2/2$. For the other two axes
lying in the plane of the disk one finds $I = I_{D}/2$. We denote
the modulus of the velocity of the centre of the disk with $v = |\vek v|$ and
its area with $A$.

For the angles given in Figure~\ref{fig_disk_sketch}, we will denote
$\omega_0 = \dot \psi$ and assume $\omega_0$ to be a constant of
motion. Our next assumption is that the disk's area always moves  
in an orientation with minimised air resistance as explained above,
i.e.\ $\vek v \perp \vek D$ and  
starts out in the $x$-$z$-plane without loss of generality.
The angle $\theta\in[0,\pi]$ does not differentiate between top- and
backspin, but only together with $\phi$.  
In order to describe everything with only a single angle,
we introduce $\vartheta\in[-\pi,+\pi]$
such that $\vartheta = \pm\theta$ with $\vartheta=+\pi/2$
corresponding to topspin and $\vartheta\ =-\pi/2$ to backspin.
Now the only non-trivial motion is the one described by $\vartheta$. 

As argued in the last section, the acceleration $\ddot\vartheta$ vanishes
for top- and backspin, with backspin the stable state. We expect
$\ddot\vartheta$ to be largest at $\vartheta = 0$ and symmetric around
this point. An easy choice to obtain such a behaviour is a force
proportional to $\cos\vartheta$. Now, there are two contributions to
$\ddot\vartheta$: the first stems from the lifting force $\vek
F^\mathrm{lift}$, which is tilting the disk. We expect the lifting
torque to be $c^\mathrm{pot}_\vartheta \rho r Av^2$ with the density
of air $\rho$ and some dimensionless constant
$c^\mathrm{pot}_\vartheta$. This is the well known formula for lift  
combined with a factor $r$ to account for the lever.
The second contribution, which reads $c_\vartheta^\mathrm{rot} \rho r
A^2 \omega_0^2$ with another constant $c^\mathrm{rot}_\vartheta$,
cannot be motivated in such a naive way. It might stem from
sub-leading effects, such as the Magnus force, but we introduced it as
an a posteriori adjustment to the empirical evidence. 

The larger $\omega_0$, the harder it will be to tilt the disk. The
corresponding force must depend on the sign of $\dot\vartheta$, but not
on the sign of $\omega_0$. Thus, we expect $c_\vartheta^\mathrm{damp}
I \omega_0^2\dot\vartheta$, which is in the form of laminar damping.
We assume $\dot\vartheta$ to be relatively small such that damping
proportional to $\dot\vartheta^2$ can be neglected.

In summary, for $\vartheta$ we arrive at the following equation of
motion with $K=\rho r A/I$
\begin{align}
  \ddot\vartheta &= -K\left(c^\text{pot}_\vartheta v^2+c^\text{rot}_\vartheta A \omega_0^2\right) \cos\vartheta - 2c^\text{damp}_\vartheta \omega_0^2 \dot\vartheta\,.\label{eqn_eff_theta_exact}
\end{align}

Next we focus on the centre of mass coordinate $\vek x = (x,y,z)$.
The aforementioned condition of minimised air resistance
prevents the disk to move in the direction of $\vek D$. This translates to an effectively reduced gravitational force $-mg\left(\hat{ \vek e}_z-\varepsilon \vek D \cos\vartheta\right)$.
The coefficient $\cos\vartheta$ takes the dependence of the air resistance in $\vek D$-direction on the projected area into account. The case $\varepsilon=0$ would correspond to a free fall, i.e.\ the leading edge would be aligned only with the horizontal velocity component and completely ignore the vertical one. Strict alignment would on the other hand imply $\varepsilon=1$. The truth, as so often, lies somewhere in between.
	
As before we introduce a damping force with the proportionality constant $c^\text{damp}_x$ of dimension velocity. Our last assumptions are $v_y=0$ at $t=0$ without loss of generality and $|v_y|\ll|v_x|$ at all times. The latter assumption is not always justified, but usually people throw disks with enough forward momentum that any change of direction is a second order effect. These assumptions allow to write $\vek D=\left(0,\sin\vartheta,\cos\vartheta\right)^\top$ significantly simplifying the equations of motion which then read
\begin{align}
	m \ddot {\vek x} &= -mg\matr{c}{0\\-\varepsilon\sin\vartheta\cos\vartheta\\1-\varepsilon\cos^2\vartheta}-c^\text{damp}_x \rho A \dot {\vek x}\,.\label{eqn_eff_x_exact}
\end{align}

\subsection{Approximate analytic solution}
Equations~\eqref{eqn_eff_theta_exact} and~\eqref{eqn_eff_x_exact} still form a set of four non-linear coupled differential equations, but they can be solved numerically much easier and more stably than the full equations of motion described in the appendix. Furthermore they can be approximately solved analytically.
	
To this end we first assume that $v\approx v_0=\text{const.}$ over the complete duration of the flight. Then Equation~\eqref{eqn_eff_theta_exact} decouples and can be solved independently of the trajectory. We are going to do so in the harmonic approximation about the potential minimum at $\vartheta=-\pi/2$. It suffices to gain a good understanding of the different stages of the flight. The differential Equation~\eqref{eqn_eff_theta_exact} simplifies to
\begin{align}
  \ddot \vartheta + 2c^\text{damp}_\vartheta \omega_0^2 \dot\vartheta + K\left(c^\text{pot}_\vartheta v_0^2+c^\text{rot}_\vartheta A \omega_0^2\right)\left(\vartheta+\frac\pi 2\right) &= 0\,.
\end{align}
We observe that the system is clearly overdamped in the case of a
stable flight (no oscillations are visible).
The corresponding stability condition for such a flight reads
\begin{align}
  \left(c^\text{damp}_\vartheta \omega_0^2\right)^2 &\overset{!}{>} K\left(c^\text{pot}_\vartheta v_0^2+c^\text{rot}_\vartheta A\omega_0^2\right)\label{eqn_stability_condition}\,.
\end{align}
In this case we obtain the well known solution
\begin{equation}
  \label{eqn_eff_sol_theta} 
  \vartheta(t) \approx -\frac\pi 2 + \left(\vartheta(0)+\frac\pi 2\right)\eto{-\lambda t}\,,
\end{equation}
with $\lambda$ given by
\[
\lambda = c^\text{damp}_\vartheta \omega_0^2 - \sqrt{\left(c^\text{damp}_\vartheta \omega_0^2\right)^2 - K\left(c^\text{pot}_\vartheta v_0^2+c^\text{rot}_\vartheta A\omega_0^2\right)}\,.
\]
In the overdamped case $\lambda$ can be simplified to
\begin{align}
  \lambda  &\approx \frac{\rho r A\left(c^\text{pot}_\vartheta v_0^2+c^\text{rot}_\vartheta A\omega_0^2\right)}{2c^\text{damp}_\vartheta I\omega_0^2}\\
	&\eqqcolon \frac{\rho r}{m}\left(\lambda_1 \frac{v_0^2}{\omega_0^2} + \lambda_0 A\right)\label{eqn_approx_sol_lambda}
\end{align}
by using $A = \pi r^2$ and $I=mr^2/4$.
Thus, the convergence depends solely on two universal constants $\lambda_0,\lambda_1$ and the four variables mass, radius, speed and angular speed. It is very remarkable that only the ratio $v_0/\omega_0$ is relevant.

With these approximations the remaining differential equations can be
solved analytically using hypergeometric functions.
However, for the sake of gaining intuitive understanding, we
approximate even further.
With sidespin initial conditions
\begin{align}
  \vartheta(0) = \dot\vartheta(0)=0\,,\quad \vek x(0)=\matr{c}{0\\0\\h}\,,\quad\dot {\vek x}(0)=\matr{c}{v_0\\0\\0}\label{eqn_default_init_conditions}
\end{align}
we focus on the $y$-component. Exploiting our condition of slow
movement in this direction allows us to neglect the damping term.
This leaves us with
\begin{align}
  \ddot y &= g\,\varepsilon\,\sin\vartheta\cos\vartheta\,.\label{eqn_eff_ode_x2}
\end{align}
Using the solution for $\vartheta$ from
Equation~\eqref{eqn_eff_sol_theta} we find that there is no
acceleration in the $y$-direction at zero and again at infinite time.
In between the acceleration becomes maximal at $\vartheta=-\pi/4$ 
with $|\ddot y|=g\varepsilon/2$ at the characteristic time scale
$\tau\coloneqq\log2/\lambda$. In fact, the acceleration is effectively
non-zero only for a short time $t_i\ll \tau$ around $\tau$ defining two
asymptotic regimes: before $\tau$ we have $\dot y=0$ and 
after $\dot y=\mathrm{const}\neq0$. The transition between the two
regimes takes place exponentially
\begin{align}
  y(t) &\approx -c_2g\varepsilon \tau t_i\log\left(1+\eto{\left(t-\tau\right)/t_i}\right)\label{eqn_eff_sol_x2}\\
  &\approx \begin{cases}
    0 & t<\tau\\
    -c_2g\varepsilon \tau\left(t-\tau\right) & t>\tau
  \end{cases}\label{eqn_eff_sol_lin_x2}
\end{align}
with some numeric prefactor $c_2$ of order one depending on
the exact form of $\vartheta$, which can be obtained by evaluating the
integral of Equation~\eqref{eqn_eff_ode_x2}.

Further details, in particular the discussions of $x$- and $z$-components, can be found in \Cref{sec_x_and_z_eff_theory}.

%% file: sections/experiments.tex
\section{Experimental method}\label{sec:experiment}

\begin{figure}[ht]
  \centering
  \includegraphics[width=0.45\textwidth]{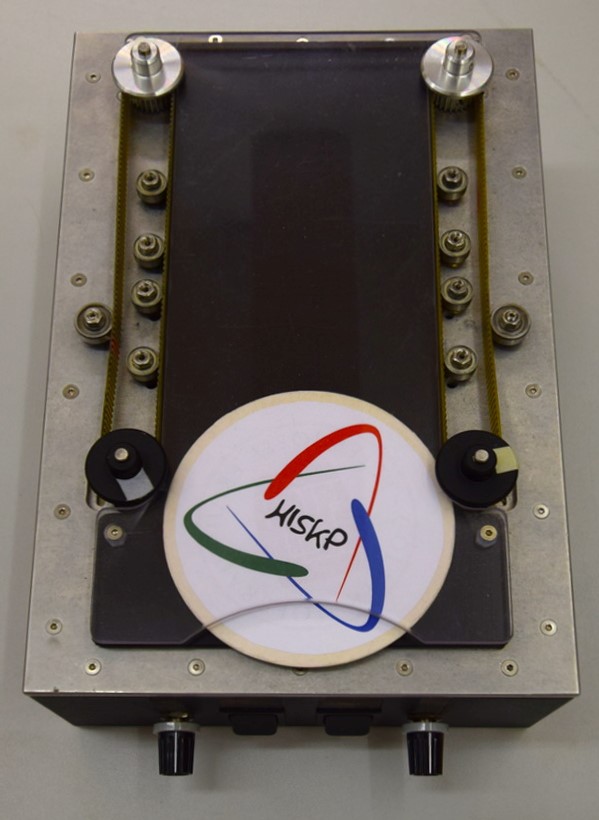}
  \caption{The beer mat shooting apparatus.}\label{fig:foto_machine}
\end{figure}

The beer mat shooting apparatus is picturised in
\Cref{fig:foto_machine}. It consists of two electric motor powered
treadmills which can be programmed to move at a given speed up to
$\SI{16}{\meter/\second}$ forwards or backwards independently of each
other. Each of the two treadmills runs around two gears with
radius $\SI{10}{\milli\meter}$, one of which (the black ones in
\Cref{fig:foto_machine}) 
is connected to the electro motor. The speed of the treadmills can be
inferred by measuring the rotations per minute of the driving gear
using a digital laser non-contact photo tachometer. We denote the
rotations per minute of the left and the right driving gear by $u_l$
and $u_r$, respectively.

A beer mat put between these treadmills is accelerated until
the edges assume the speed of the respective treadmill. We have
checked with high speed camera videos that an undamaged beer mat does
indeed not slip along the treadmills. During acceleration the beer mat
is confined vertically between two plastic surfaces.

During operation the apparatus is placed on a table so that the beer
mats are shot from a height of $\SI{0.98}{\meter}$ with the plastic
surfaces horizontally aligned. Then, to the level of accuracy required here the 
apparatus produces reproducible initial flight conditions. All beer
mats shot with the same treadmill configuration followed the same
trajectory and hit the same point on the floor, up to minor deviations
in the order of $\SI{0.1}{\meter}$.

Flights were recorded using a high speed camera\footnote{CR600x2, Optronis Slow-Motion Camera} with
$500$ frames per second.
Example videos of flights are available in the supplemental material.
We used the program Tracker\footnote{\url{https://physlets.org/tracker/}} to extract the coordinates of the beer mat at any given time from the videos.
The measurements cover a broad range of different initial velocities and angular momenta.
A summary of the recorded and analysed experimental setups is provided in table~\ref{tab:sets} of the appendix.


%% file: sections/results.tex
\begin{figure}[t]
  \centering
  \input{figures/y_t_5000_zu_0_hinten}
  \caption{Horizontal position $y$ orthogonal to the initial flight
    direction in meters against time in seconds. The flight started
    with initial conditions as in
    Eq.~\eqref{eqn_default_init_conditions} with
    $v_0=\SI{2.6}{\meter/\second}$ and
    $\omega_0=\SI{49}{\second}^{-1}$ (corresponding to $\num{7.9}$
    rotations per second). The effective theory fit follows
    Eq.~\eqref{eqn_eff_sol_x2}. For details on the numerical
    simulation see \cref{sec:numerical_simulation}. The lengths are not
    exactly correct due to perspective
    distortion.}\label{fig_y_t_trajectory} 
\end{figure}
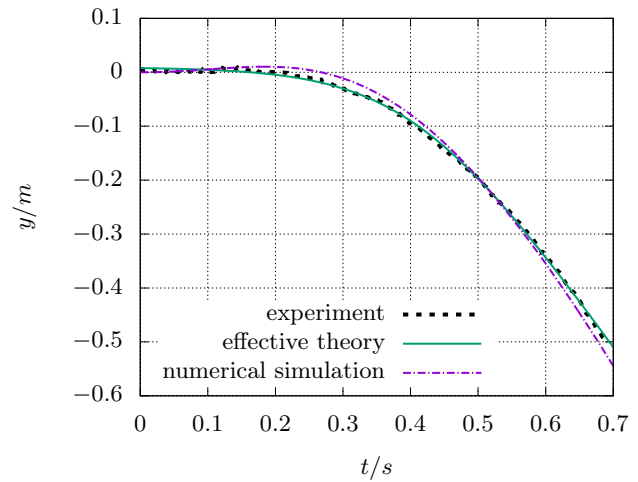

\section{Results and Discussion}

For each of the recorded experiments we have analysed the trajectory
of the corresponding flight. First, we observe that the angular
velocity $\omega_0$ changes by less than $10\%$ during the flights.

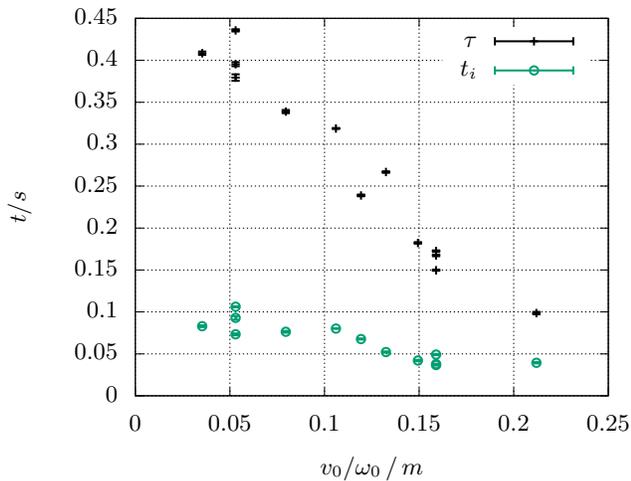
\begin{figure}[t]
  \centering
  \input{figures/tis}
  \caption{Characteristic times $t_i$ and $\tau$ derived via a fit of
    Eq.~\eqref{eqn_eff_sol_x2} for a variety of different initial
    conditions in speed $v_0$ and angular speed
    $\omega_0$.}\label{fig_characteristic_times} 
\end{figure}

Next we reconstruct the horizontal position $y(t)$ from the
recorded video. We fit the functional form
Equation~\eqref{eqn_eff_sol_x2} to this data for $y(t)$ for each 
experimental setup separately and determine the parameters $c_2\epsilon,
t_i$ and $\tau$. One representative example of such a
$y$-trajectory and the corresponding fit are shown in
Figure~\ref{fig_y_t_trajectory}. The so determined values for the
characteristic time $\tau$ and $t_i$ are shown in
Figure~\ref{fig_characteristic_times} as functions of
$v_0/\omega_0$. From this Figure one can already read off that stable
flights of beer mats longer than $\SI{0.45}{\second}$ are hardly
possible. 
The detailed fit results are provided in table~\ref{tab:fit_results} of the appendix.

Note that we also solved the full set of differential equations
including lift, drag and sub-leading effects numerically, for details
see \cref{sec:numerical_simulation}. The corresponding $y$-trajectory is also depicted in
\Cref{fig_y_t_trajectory} for means of comparison as the purple dash-dotted
line. 

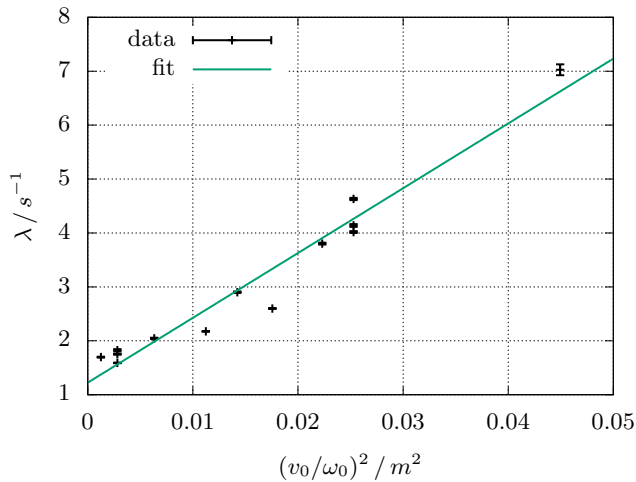
\begin{figure}[t]
  \centering
  \input{figures/lambdas}
  \caption{Damping factor $\lambda$ derived via a fit of
    Eq.~\eqref{eqn_eff_sol_x2} for a variety of different initial
    conditions in speed $v_0$ and angular speed $\omega_0$. The fit
    follows Eq.~\eqref{eqn_approx_sol_lambda}.}\label{fig_lambdas} 
\end{figure}

In Figure~\ref{fig_lambdas} we show $\lambda$ determined via $\lambda
= \log2/\tau$ as a function of $(v_0/\omega_0)^2$. In the same Figure
we show a fit of Eq.~\eqref{eqn_approx_sol_lambda} to the data for
$\lambda$ (with
$\rho=\SI{1.25}{\kilogram/\meter\cubed}$, $m=\SI{5.9}{\gram}$ and
$A=\pi r^2$, $r=\SI{5.3}{\centi\meter}$) yielding $\lambda_0 =
\SI{12.4\pm 1.5}{\second}^{-1}$ and $\lambda_1 = \SI{10.7\pm
  0.7}{\second}^{-1}$.

Note that the statistical error from the fit depicted with errorbars
has but a small contribution to the overall error. The total error
comes mostly from the fact that the flight of a beer mat is not
perfectly reproducible with our apparatus and in principal identical
initial conditions lead to slightly different results when repeated
several times. This discrepancy between statistical and total error is
reflected in the large spread of points totally incompatible with the
size of their error bars.

As we mentioned above, $\omega_0$ does change measurably during the
flight, however, not significantly in particular in the phase of the
flight relevant for our analysis. Therefore, we think the main
assumption for our effective theory of constant $\omega_0$ is
fulfilled well enough, even more so as the effective theory, namely
Equation~\eqref{eqn_eff_sol_x2} (and~\eqref{eqn_eff_sol_x1},
\eqref{eqn_eff_sol_x3} from the appendix), describes the
flight trajectories extremely well with fitted parameters. 
We find our earlier expectations confirmed as indeed $t_i\ll \tau$ for
all flights we investigated. Most importantly, the values of $\lambda$
are in good agreement with a linear dependence on $v_0^2/\omega_0^2$
plus a constant. 


These values for $\lambda_0$ and $\lambda_1$ are universal constants
for every system describable by the effective theory. Thus, they
should allow us to predict the duration of a stable flight of any
rotating and thin enough disk. Just using the constant term
$\propto\lambda_0$ solely  
depending on the disk's properties allows one to predict the longest
possible stable flight duration. We arrive
at some very interesting and astonishingly realistic
predictions: recall that a beer mat maintains stability for up to 
about $\SI{0.45}{\second}$. Now a standard playing card\footnote{Even
  Rick Smith, Jr., the world record holder for farthest card thrown,
  or a playing card machine gun cannot avoid their cards flying a
  curve and ending up with backspin after much less than a
  second~\cite{playing_cards}.} is expected to reach slightly more
than half that time, namely about $\SI{0.24}{\second}$, a CD holds out
for twice a beer mat's time or $\sim\SI{0.8}{\second}$, a frisbee for
ca.~$\SI{0.7}{\second}$ (without any aerodynamics due to curvature
taken into account) and a discus could in principle fly undisturbed
for up to ca.~$\SI{16}{\second}$. These predictions could be tested
experimentally. 

We find especially the time estimate for the frisbee very interesting
because it is not unrealistic for a mediocre thrower to have the
frisbee flip towards backspin in the timespan of about one second
though this estimate lies rather on the short side and of course the
wing form of a frisbee allows it to remain stable for a much longer
time when thrown professionally. The reason is that frisbees have
their aerodynamic center very near to their center of
mass~\cite{frisbee_thesis} and, thus, experience much less torque.

%% file: figures/y_t_5000_zu_0_hinten.tex
\begingroup
  \inputencoding{latin1}%
  \makeatletter
  \providecommand\color[2][]{%
    \GenericError{(gnuplot) \space\space\space\@spaces}{%
      Package color not loaded in conjunction with
      terminal option `colourtext'%
    }{See the gnuplot documentation for explanation.%
    }{Either use 'blacktext' in gnuplot or load the package
      color.sty in LaTeX.}%
    \renewcommand\color[2][]{}%
  }%
  \providecommand\includegraphics[2][]{%
    \GenericError{(gnuplot) \space\space\space\@spaces}{%
      Package graphicx or graphics not loaded%
    }{See the gnuplot documentation for explanation.%
    }{The gnuplot epslatex terminal needs graphicx.sty or graphics.sty.}%
    \renewcommand\includegraphics[2][]{}%
  }%
  \providecommand\rotatebox[2]{#2}%
  \@ifundefined{ifGPcolor}{%
    \newif\ifGPcolor
    \GPcolortrue
  }{}%
  \@ifundefined{ifGPblacktext}{%
    \newif\ifGPblacktext
    \GPblacktexttrue
  }{}%
  \let\gplgaddtomacro\g@addto@macro
  \gdef\gplbacktext{}%
  \gdef\gplfronttext{}%
  \makeatother
  \ifGPblacktext
    \def\colorrgb#1{}%
    \def\colorgray#1{}%
  \else
    \ifGPcolor
      \def\colorrgb#1{\color[rgb]{#1}}%
      \def\colorgray#1{\color[gray]{#1}}%
      \expandafter\def\csname LTw\endcsname{\color{white}}%
      \expandafter\def\csname LTb\endcsname{\color{black}}%
      \expandafter\def\csname LTa\endcsname{\color{black}}%
      \expandafter\def\csname LT0\endcsname{\color[rgb]{1,0,0}}%
      \expandafter\def\csname LT1\endcsname{\color[rgb]{0,1,0}}%
      \expandafter\def\csname LT2\endcsname{\color[rgb]{0,0,1}}%
      \expandafter\def\csname LT3\endcsname{\color[rgb]{1,0,1}}%
      \expandafter\def\csname LT4\endcsname{\color[rgb]{0,1,1}}%
      \expandafter\def\csname LT5\endcsname{\color[rgb]{1,1,0}}%
      \expandafter\def\csname LT6\endcsname{\color[rgb]{0,0,0}}%
      \expandafter\def\csname LT7\endcsname{\color[rgb]{1,0.3,0}}%
      \expandafter\def\csname LT8\endcsname{\color[rgb]{0.5,0.5,0.5}}%
    \else
      \def\colorrgb#1{\color{black}}%
      \def\colorgray#1{\color[gray]{#1}}%
      \expandafter\def\csname LTw\endcsname{\color{white}}%
      \expandafter\def\csname LTb\endcsname{\color{black}}%
      \expandafter\def\csname LTa\endcsname{\color{black}}%
      \expandafter\def\csname LT0\endcsname{\color{black}}%
      \expandafter\def\csname LT1\endcsname{\color{black}}%
      \expandafter\def\csname LT2\endcsname{\color{black}}%
      \expandafter\def\csname LT3\endcsname{\color{black}}%
      \expandafter\def\csname LT4\endcsname{\color{black}}%
      \expandafter\def\csname LT5\endcsname{\color{black}}%
      \expandafter\def\csname LT6\endcsname{\color{black}}%
      \expandafter\def\csname LT7\endcsname{\color{black}}%
      \expandafter\def\csname LT8\endcsname{\color{black}}%
    \fi
  \fi
    \setlength{\unitlength}{0.0500bp}%
    \ifx\gptboxheight\undefined%
      \newlength{\gptboxheight}%
      \newlength{\gptboxwidth}%
      \newsavebox{\gptboxtext}%
    \fi%
    \setlength{\fboxrule}{0.5pt}%
    \setlength{\fboxsep}{1pt}%
\begin{picture}(5040.00,3772.00)%
    \gplgaddtomacro\gplbacktext{%
      \csname LTb\endcsname
      \put(946,704){\makebox(0,0)[r]{\strut{}$-0.6$}}%
      \csname LTb\endcsname
      \put(946,1111){\makebox(0,0)[r]{\strut{}$-0.5$}}%
      \csname LTb\endcsname
      \put(946,1517){\makebox(0,0)[r]{\strut{}$-0.4$}}%
      \csname LTb\endcsname
      \put(946,1924){\makebox(0,0)[r]{\strut{}$-0.3$}}%
      \csname LTb\endcsname
      \put(946,2331){\makebox(0,0)[r]{\strut{}$-0.2$}}%
      \csname LTb\endcsname
      \put(946,2738){\makebox(0,0)[r]{\strut{}$-0.1$}}%
      \csname LTb\endcsname
      \put(946,3144){\makebox(0,0)[r]{\strut{}$0$}}%
      \csname LTb\endcsname
      \put(946,3551){\makebox(0,0)[r]{\strut{}$0.1$}}%
      \csname LTb\endcsname
      \put(1078,484){\makebox(0,0){\strut{}$0$}}%
      \csname LTb\endcsname
      \put(1587,484){\makebox(0,0){\strut{}$0.1$}}%
      \csname LTb\endcsname
      \put(2097,484){\makebox(0,0){\strut{}$0.2$}}%
      \csname LTb\endcsname
      \put(2606,484){\makebox(0,0){\strut{}$0.3$}}%
      \csname LTb\endcsname
      \put(3115,484){\makebox(0,0){\strut{}$0.4$}}%
      \csname LTb\endcsname
      \put(3624,484){\makebox(0,0){\strut{}$0.5$}}%
      \csname LTb\endcsname
      \put(4134,484){\makebox(0,0){\strut{}$0.6$}}%
      \csname LTb\endcsname
      \put(4643,484){\makebox(0,0){\strut{}$0.7$}}%
    }%
    \gplgaddtomacro\gplfronttext{%
      \csname LTb\endcsname
      \put(209,2127){\rotatebox{-270}{\makebox(0,0){\strut{}$y/m$}}}%
      \put(2860,154){\makebox(0,0){\strut{}$t/s$}}%
      \csname LTb\endcsname
      \put(2926,1317){\makebox(0,0)[r]{\strut{}experiment}}%
      \csname LTb\endcsname
      \put(2926,1097){\makebox(0,0)[r]{\strut{}effective theory}}%
      \csname LTb\endcsname
      \put(2926,877){\makebox(0,0)[r]{\strut{}numerical simulation}}%
    }%
    \gplbacktext
    \put(0,0){\includegraphics{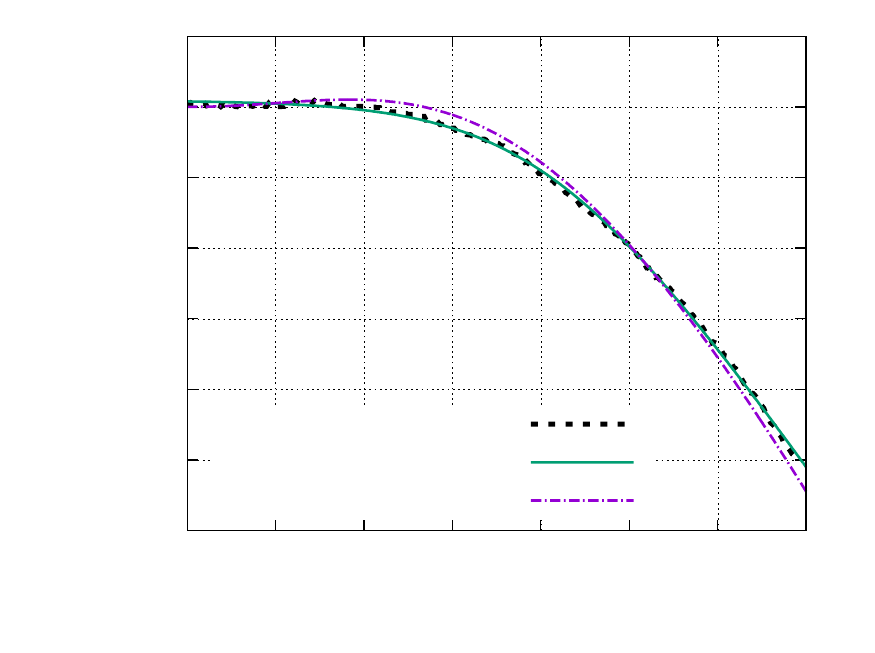}}%
    \gplfronttext
  \end{picture}%
\endgroup

%% file: figures/tis.tex
\begingroup
  \inputencoding{latin1}%
  \makeatletter
  \providecommand\color[2][]{%
    \GenericError{(gnuplot) \space\space\space\@spaces}{%
      Package color not loaded in conjunction with
      terminal option `colourtext'%
    }{See the gnuplot documentation for explanation.%
    }{Either use 'blacktext' in gnuplot or load the package
      color.sty in LaTeX.}%
    \renewcommand\color[2][]{}%
  }%
  \providecommand\includegraphics[2][]{%
    \GenericError{(gnuplot) \space\space\space\@spaces}{%
      Package graphicx or graphics not loaded%
    }{See the gnuplot documentation for explanation.%
    }{The gnuplot epslatex terminal needs graphicx.sty or graphics.sty.}%
    \renewcommand\includegraphics[2][]{}%
  }%
  \providecommand\rotatebox[2]{#2}%
  \@ifundefined{ifGPcolor}{%
    \newif\ifGPcolor
    \GPcolortrue
  }{}%
  \@ifundefined{ifGPblacktext}{%
    \newif\ifGPblacktext
    \GPblacktexttrue
  }{}%
  \let\gplgaddtomacro\g@addto@macro
  \gdef\gplbacktext{}%
  \gdef\gplfronttext{}%
  \makeatother
  \ifGPblacktext
    \def\colorrgb#1{}%
    \def\colorgray#1{}%
  \else
    \ifGPcolor
      \def\colorrgb#1{\color[rgb]{#1}}%
      \def\colorgray#1{\color[gray]{#1}}%
      \expandafter\def\csname LTw\endcsname{\color{white}}%
      \expandafter\def\csname LTb\endcsname{\color{black}}%
      \expandafter\def\csname LTa\endcsname{\color{black}}%
      \expandafter\def\csname LT0\endcsname{\color[rgb]{1,0,0}}%
      \expandafter\def\csname LT1\endcsname{\color[rgb]{0,1,0}}%
      \expandafter\def\csname LT2\endcsname{\color[rgb]{0,0,1}}%
      \expandafter\def\csname LT3\endcsname{\color[rgb]{1,0,1}}%
      \expandafter\def\csname LT4\endcsname{\color[rgb]{0,1,1}}%
      \expandafter\def\csname LT5\endcsname{\color[rgb]{1,1,0}}%
      \expandafter\def\csname LT6\endcsname{\color[rgb]{0,0,0}}%
      \expandafter\def\csname LT7\endcsname{\color[rgb]{1,0.3,0}}%
      \expandafter\def\csname LT8\endcsname{\color[rgb]{0.5,0.5,0.5}}%
    \else
      \def\colorrgb#1{\color{black}}%
      \def\colorgray#1{\color[gray]{#1}}%
      \expandafter\def\csname LTw\endcsname{\color{white}}%
      \expandafter\def\csname LTb\endcsname{\color{black}}%
      \expandafter\def\csname LTa\endcsname{\color{black}}%
      \expandafter\def\csname LT0\endcsname{\color{black}}%
      \expandafter\def\csname LT1\endcsname{\color{black}}%
      \expandafter\def\csname LT2\endcsname{\color{black}}%
      \expandafter\def\csname LT3\endcsname{\color{black}}%
      \expandafter\def\csname LT4\endcsname{\color{black}}%
      \expandafter\def\csname LT5\endcsname{\color{black}}%
      \expandafter\def\csname LT6\endcsname{\color{black}}%
      \expandafter\def\csname LT7\endcsname{\color{black}}%
      \expandafter\def\csname LT8\endcsname{\color{black}}%
    \fi
  \fi
    \setlength{\unitlength}{0.0500bp}%
    \ifx\gptboxheight\undefined%
      \newlength{\gptboxheight}%
      \newlength{\gptboxwidth}%
      \newsavebox{\gptboxtext}%
    \fi%
    \setlength{\fboxrule}{0.5pt}%
    \setlength{\fboxsep}{1pt}%
\begin{picture}(5040.00,3772.00)%
    \gplgaddtomacro\gplbacktext{%
      \csname LTb\endcsname
      \put(946,704){\makebox(0,0)[r]{\strut{}$0$}}%
      \csname LTb\endcsname
      \put(946,1020){\makebox(0,0)[r]{\strut{}$0.05$}}%
      \csname LTb\endcsname
      \put(946,1337){\makebox(0,0)[r]{\strut{}$0.1$}}%
      \csname LTb\endcsname
      \put(946,1653){\makebox(0,0)[r]{\strut{}$0.15$}}%
      \csname LTb\endcsname
      \put(946,1969){\makebox(0,0)[r]{\strut{}$0.2$}}%
      \csname LTb\endcsname
      \put(946,2286){\makebox(0,0)[r]{\strut{}$0.25$}}%
      \csname LTb\endcsname
      \put(946,2602){\makebox(0,0)[r]{\strut{}$0.3$}}%
      \csname LTb\endcsname
      \put(946,2918){\makebox(0,0)[r]{\strut{}$0.35$}}%
      \csname LTb\endcsname
      \put(946,3235){\makebox(0,0)[r]{\strut{}$0.4$}}%
      \csname LTb\endcsname
      \put(946,3551){\makebox(0,0)[r]{\strut{}$0.45$}}%
      \csname LTb\endcsname
      \put(1078,484){\makebox(0,0){\strut{}$0$}}%
      \csname LTb\endcsname
      \put(1791,484){\makebox(0,0){\strut{}$0.05$}}%
      \csname LTb\endcsname
      \put(2504,484){\makebox(0,0){\strut{}$0.1$}}%
      \csname LTb\endcsname
      \put(3217,484){\makebox(0,0){\strut{}$0.15$}}%
      \csname LTb\endcsname
      \put(3930,484){\makebox(0,0){\strut{}$0.2$}}%
      \csname LTb\endcsname
      \put(4643,484){\makebox(0,0){\strut{}$0.25$}}%
    }%
    \gplgaddtomacro\gplfronttext{%
      \csname LTb\endcsname
      \put(209,2127){\rotatebox{-270}{\makebox(0,0){\strut{}$t/s$}}}%
      \put(2860,154){\makebox(0,0){\strut{}$v_0/\omega_0\, /\, m$}}%
      \csname LTb\endcsname
      \put(3656,3378){\makebox(0,0)[r]{\strut{}$\tau$}}%
      \csname LTb\endcsname
      \put(3656,3158){\makebox(0,0)[r]{\strut{}$t_i$}}%
    }%
    \gplbacktext
    \put(0,0){\includegraphics{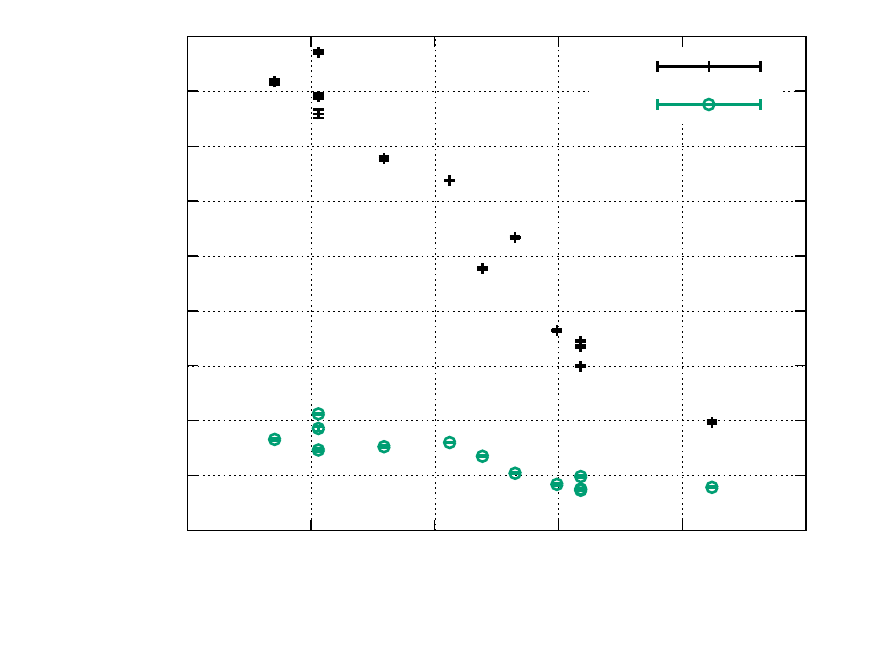}}%
    \gplfronttext
  \end{picture}%
\endgroup

%% file: figures/lambdas.tex
\begingroup
  \inputencoding{latin1}%
  \makeatletter
  \providecommand\color[2][]{%
    \GenericError{(gnuplot) \space\space\space\@spaces}{%
      Package color not loaded in conjunction with
      terminal option `colourtext'%
    }{See the gnuplot documentation for explanation.%
    }{Either use 'blacktext' in gnuplot or load the package
      color.sty in LaTeX.}%
    \renewcommand\color[2][]{}%
  }%
  \providecommand\includegraphics[2][]{%
    \GenericError{(gnuplot) \space\space\space\@spaces}{%
      Package graphicx or graphics not loaded%
    }{See the gnuplot documentation for explanation.%
    }{The gnuplot epslatex terminal needs graphicx.sty or graphics.sty.}%
    \renewcommand\includegraphics[2][]{}%
  }%
  \providecommand\rotatebox[2]{#2}%
  \@ifundefined{ifGPcolor}{%
    \newif\ifGPcolor
    \GPcolortrue
  }{}%
  \@ifundefined{ifGPblacktext}{%
    \newif\ifGPblacktext
    \GPblacktexttrue
  }{}%
  \let\gplgaddtomacro\g@addto@macro
  \gdef\gplbacktext{}%
  \gdef\gplfronttext{}%
  \makeatother
  \ifGPblacktext
    \def\colorrgb#1{}%
    \def\colorgray#1{}%
  \else
    \ifGPcolor
      \def\colorrgb#1{\color[rgb]{#1}}%
      \def\colorgray#1{\color[gray]{#1}}%
      \expandafter\def\csname LTw\endcsname{\color{white}}%
      \expandafter\def\csname LTb\endcsname{\color{black}}%
      \expandafter\def\csname LTa\endcsname{\color{black}}%
      \expandafter\def\csname LT0\endcsname{\color[rgb]{1,0,0}}%
      \expandafter\def\csname LT1\endcsname{\color[rgb]{0,1,0}}%
      \expandafter\def\csname LT2\endcsname{\color[rgb]{0,0,1}}%
      \expandafter\def\csname LT3\endcsname{\color[rgb]{1,0,1}}%
      \expandafter\def\csname LT4\endcsname{\color[rgb]{0,1,1}}%
      \expandafter\def\csname LT5\endcsname{\color[rgb]{1,1,0}}%
      \expandafter\def\csname LT6\endcsname{\color[rgb]{0,0,0}}%
      \expandafter\def\csname LT7\endcsname{\color[rgb]{1,0.3,0}}%
      \expandafter\def\csname LT8\endcsname{\color[rgb]{0.5,0.5,0.5}}%
    \else
      \def\colorrgb#1{\color{black}}%
      \def\colorgray#1{\color[gray]{#1}}%
      \expandafter\def\csname LTw\endcsname{\color{white}}%
      \expandafter\def\csname LTb\endcsname{\color{black}}%
      \expandafter\def\csname LTa\endcsname{\color{black}}%
      \expandafter\def\csname LT0\endcsname{\color{black}}%
      \expandafter\def\csname LT1\endcsname{\color{black}}%
      \expandafter\def\csname LT2\endcsname{\color{black}}%
      \expandafter\def\csname LT3\endcsname{\color{black}}%
      \expandafter\def\csname LT4\endcsname{\color{black}}%
      \expandafter\def\csname LT5\endcsname{\color{black}}%
      \expandafter\def\csname LT6\endcsname{\color{black}}%
      \expandafter\def\csname LT7\endcsname{\color{black}}%
      \expandafter\def\csname LT8\endcsname{\color{black}}%
    \fi
  \fi
    \setlength{\unitlength}{0.0500bp}%
    \ifx\gptboxheight\undefined%
      \newlength{\gptboxheight}%
      \newlength{\gptboxwidth}%
      \newsavebox{\gptboxtext}%
    \fi%
    \setlength{\fboxrule}{0.5pt}%
    \setlength{\fboxsep}{1pt}%
\begin{picture}(5040.00,3772.00)%
    \gplgaddtomacro\gplbacktext{%
      \csname LTb\endcsname
      \put(550,704){\makebox(0,0)[r]{\strut{}$1$}}%
      \csname LTb\endcsname
      \put(550,1111){\makebox(0,0)[r]{\strut{}$2$}}%
      \csname LTb\endcsname
      \put(550,1517){\makebox(0,0)[r]{\strut{}$3$}}%
      \csname LTb\endcsname
      \put(550,1924){\makebox(0,0)[r]{\strut{}$4$}}%
      \csname LTb\endcsname
      \put(550,2331){\makebox(0,0)[r]{\strut{}$5$}}%
      \csname LTb\endcsname
      \put(550,2738){\makebox(0,0)[r]{\strut{}$6$}}%
      \csname LTb\endcsname
      \put(550,3144){\makebox(0,0)[r]{\strut{}$7$}}%
      \csname LTb\endcsname
      \put(550,3551){\makebox(0,0)[r]{\strut{}$8$}}%
      \csname LTb\endcsname
      \put(682,484){\makebox(0,0){\strut{}$0$}}%
      \csname LTb\endcsname
      \put(1474,484){\makebox(0,0){\strut{}$0.01$}}%
      \csname LTb\endcsname
      \put(2266,484){\makebox(0,0){\strut{}$0.02$}}%
      \csname LTb\endcsname
      \put(3059,484){\makebox(0,0){\strut{}$0.03$}}%
      \csname LTb\endcsname
      \put(3851,484){\makebox(0,0){\strut{}$0.04$}}%
      \csname LTb\endcsname
      \put(4643,484){\makebox(0,0){\strut{}$0.05$}}%
    }%
    \gplgaddtomacro\gplfronttext{%
      \csname LTb\endcsname
      \put(209,2127){\rotatebox{-270}{\makebox(0,0){\strut{}$\lambda\, /\, s^{-1}$}}}%
      \put(2662,154){\makebox(0,0){\strut{}$(v_0/\omega_0)^2\, /\, m^2$}}%
      \csname LTb\endcsname
      \put(1342,3378){\makebox(0,0)[r]{\strut{}data}}%
      \csname LTb\endcsname
      \put(1342,3158){\makebox(0,0)[r]{\strut{}fit}}%
    }%
    \gplbacktext
    \put(0,0){\includegraphics{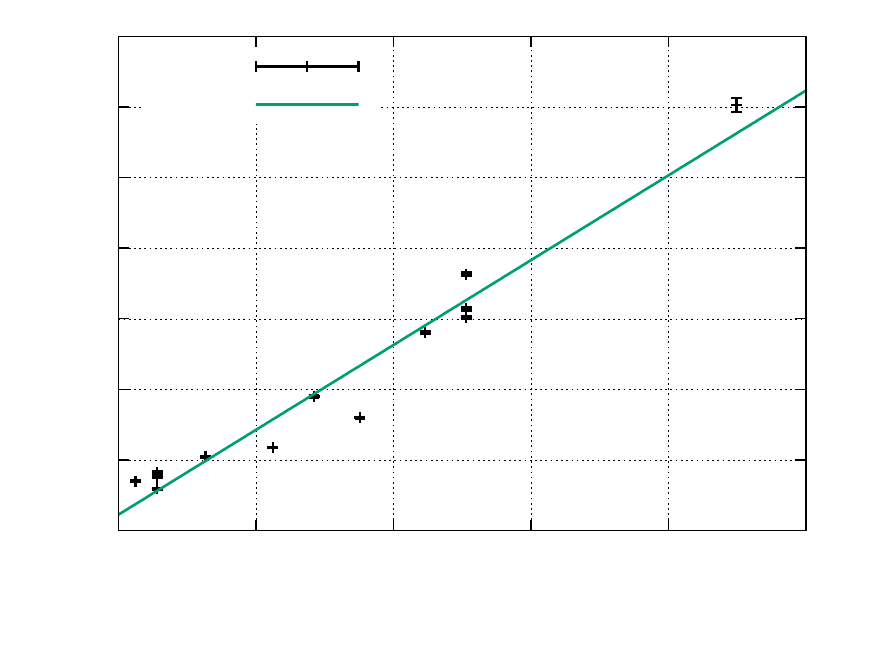}}%
    \gplfronttext
  \end{picture}%
\endgroup

%% file: sections/conclusion.tex
\section{Summary}

In this paper we have investigated the peculiar flight trajectories of
beer mats: independently of their initial conditions, beer mats
always tilt into backspin position shortly after being thrown.
When thrown by hand, the moment this tilting starts is seemingly
random.

We have presented an explanation for this effect and developed
an effective theory describing the
flight of beer mats, or any thin disk, alongside an experimental
investigation of such flights. We used the experimental results to
estimate the parameters of the aforementioned effective theory, which
describes the data very well. The effective theory then makes
universal predictions: for instance, the damping factor $\lambda$ of
the tilting motion must depend linearly on $(v_0/\omega_0)^2$ (the
centre-of-mass velocity and the angular speed, respectively) and
otherwise only on the disk's radius, mass and the density of air.
This is nicely confirmed by our empirical results. 

$\lambda$ is directly related to the time $\tau$ via
$\lambda=\log(2)/\tau$. $\tau$ is the time after take off at which the 
tilting into backspin orientation becomes visible, i.e.\ $\tau$ is the
time of stable flight. Its apparent randomness solely stems from the
inability to reproduce initial conditions when throwing by hand.

Since the effective theory holds for any thin disk we are also able to
predict $\tau$ for other types of disks like playing cards or
CDs, which could be tested experimentally. Frisbees, however, do have
different aerodynamic properties than beer mats due to their rounded
down edges and, thus, enjoy a significantly extended stable flight
time.

%% file: sections/forces.tex
\section{Forces}\label{sec:forces}
	The equations of motion (EOM) of a rotating rigid body can be derived with well known techniques. This derivation is given in \cref{app:eom} with the variable convention as in \cref{app:convention}. Let us then get our hands on the constraining forces $F_q$ describing the non-free motion of the disk through air. The challenge is to find forms that are easy (we do not want for example to solve the complete Navier–Stokes equations) and at the same time accurate enough. There are many effective formulae for drag and lift~\cite{aerodynamik} but all of them are only approximations and it is neither clear how well they describe the given problem, nor does one know a priori which are the most significant and which can be neglected. Our numerical simulations helped to identify the most relevant terms as presented here.
	
\begin{figure}[t]
		\centering
		\includegraphics[trim={1.7cm 1.2cm 1.3cm 1.2cm},clip, width=\columnwidth]{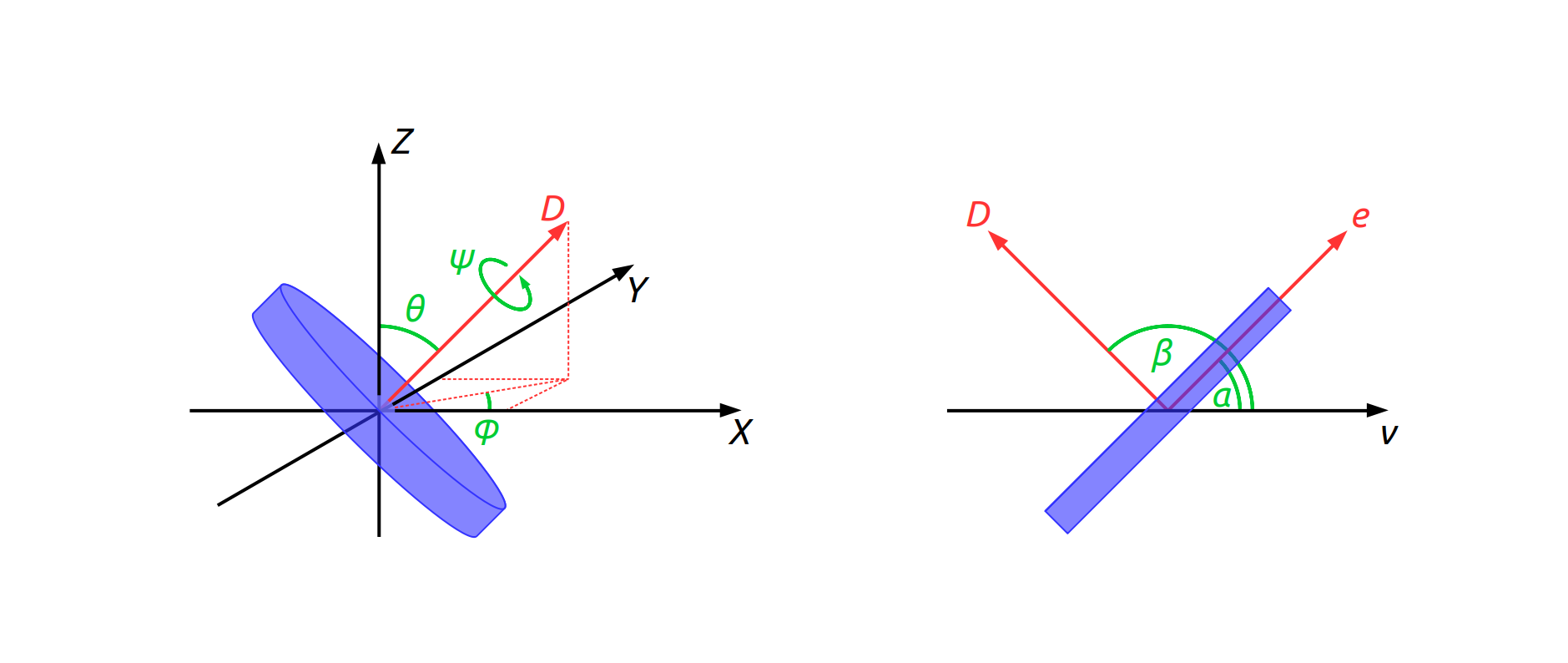}
		\caption{Sketch of the disk and the most relevant coordinates.}\label{fig_disk_sketch_both_sides}
	\end{figure}

	The most relevant coordinates are visualised in figure~\ref{fig_disk_sketch_both_sides}. A detailed summary of all the coordinates is provided in \cref{app:convention}.

	In this section we will use the notation
	\begin{equation}
	F_q \equiv \sum_i F_q^i
	\end{equation}
	where $i$ runs over all indices relevant for this component (e.g. $F_x=F_x^\text{drag}+F_x^\text{lift}$).
	
	\subsection{Derivation of the equations of motion}\label{app:eom}
	In what follows we use Lagrange's formalism because we find it clearer than Newton's laws or Euler's equations, though (of course) either approach eventually leads to the same EOM.
	
	The moment of inertia for a flat disk of mass $m$ and radius $r$ is given by
	\begin{equation}
		I = \frac14 mr^2
	\end{equation}
	about an axis lying in the plain of the disk and by $2I$ about the symmetry axis. Thus the total rotation energy is
	\begin{align}
		E_{\text{rot}}&=\frac12 I\dot{\vek\Phi}^t\matr{rrr}{1&0&0\\0&1&0\\0&0&2}\dot{\vek\Phi}\\
		&=\frac12 I\left(\dtheta^2+\left(1+\cos^2\theta\right)\dphi^2+4\ct\,\dphi\,\dpsi+2\dpsi^2\right)\,.
	\end{align}
	Here
	\begin{equation}
		\dot{\vek\Phi}=\left(\!\begin{array}{c}-\sin\phi\\\cos\phi\\0\end{array}\!\right)\dot{\theta}+\left(\!\begin{array}{c}\sin\theta\cos\phi\\\sin\theta\sin\phi\\\cos\theta\end{array}\!\right)\dphi+\left(\!\begin{array}{c}0\\0\\1\end{array}\!\right)\dot{\psi}
	\end{equation}
	denotes the total angular velocity in a basis with diagonal inertia tensor.
	
	The kinetic and potential energies are simply given by $E_{\text{kin}}=\frac12 m\dot{\vek x}^2$ and \mbox{$E_{\text{pot}}=mgz$} respectively. In addition we introduced a torque $E_{\text{tor}}=\vek T\cdot\vek D$. In the case of a flying disk we have $\vek T=0$. A non-zero torque however leads to well known precession movements and is thus very helpful for consistency checks. It corresponds to a preferred symmetry axis or a weight attached to the disk on one side.
	
	Putting all these terms together yields the Lagrange function
	\begin{align}
		L&=E_{\text{kin}}+E_{\text{rot}}-E_{\text{pot}}-E_{\text{tor}}\\
		&=\frac12 m\dot{\vek x}^2+\frac12 I\left(\dtheta^2+\left(1+\cos^2\theta\right)\dphi^2+4\ct\,\dphi\,\dpsi+2\dpsi^2\right)-mgz-\vek T\cdot\vek D\,.
	\end{align}
	For a generalised coordinate $q$ exerted to an external constraint force $F_q$ the Euler-Lagrange-equation reads
	\begin{equation}
		F_q=\frac{\md}{\md t}\left(\del{L}{\dot{q}}\right)-\del{L}{q}\,.
	\end{equation}
	As we do not know the drag and lift forces a priori we will denote them by $F_{\vek x,\theta,\phi,\psi}$ respectively. This leads to
	\begin{align}
		\begin{split}
			F_{\vek x}&=m\ddot{\vek x}+mg\matr{c}{0\\0\\1}
		\end{split}\\[2ex]
		\begin{split}
			F_\theta&=I\ddot{\theta}+I\ct\st\,\dphi^2+2I\st\,\dphi\,\dpsi+\vek T\cdot\vek D_\theta
		\end{split}\\[2ex]
		\begin{split}\label{eqn_ele_phi}
			F_\phi&= I\left(\left(1+\cos^2\theta\right)\ddot{\phi}-2\ct\st\,\dtheta\,\dphi+2\ct\ddot{\psi}-2\st\,\dtheta\,\dpsi\right)+\st\,\vek T\cdot\vek D_\phi
		\end{split}\\[2ex]
		\begin{split}\label{eqn_ele_psi}
			F_\psi&=2I\left(\ct\,\ddot{\phi}-\st\,\dtheta\,\dphi+\ddot{\psi}\right)
		\end{split}
	\end{align}
	where we defined
	\begin{equation}
		\del{\vek D}{\theta}\equiv \vek D_\theta\quad \text{and}\quad\del{\vek D}{\phi}\equiv \st\,\vek D_\phi
	\end{equation}
	using the orthogonal unit vectors in direction of increasing $\theta$ and $\phi$. In the canonical representation this means
	\begin{equation}
		\vek D=\matr{c}{\st\cp\\\st\sp\\\ct}\;,\quad \vek D_\theta=\matr{c}{\ct\cp\\\ct\sp\\-\st}\;,\quad \vek D_\phi=\matr{c}{-\sp\\\cp\\0}\;.\label{eqn_def_d}
	\end{equation}
	In order to simplify further calculations we introduce the reduced quantities corresponding to the respective accelerations
	\begin{align}
		\vek f_{\vek x}&\coloneqq \frac1m \vek F_{\vek x}\\
		f_\theta&\coloneqq \frac1I \left(F_\theta-\vek T\cdot\vek D_\theta\right)\\
		f_\phi&\coloneqq \frac1I \left(F_\phi-\st\, \vek T\cdot\vek D_\phi\right)\\
		f_\psi&\coloneqq \frac 1I F_\psi
	\end{align}
	instead of forces. With these terms and the solution of equations~\eqref{eqn_ele_phi} and~\eqref{eqn_ele_psi} for $\ddot{\phi}$ and $\ddot{\psi}$ the EOMs read
	\begin{align}
		\dot{\vek x}&=\vek v\\
		\dtheta&=\ptheta\\
		\dphi&=\pphi\\
		\dpsi&=\ppsi\\
		\dot{\vek v}&=-g\matr{c}{0\\0\\1}+\vek f_{\vek x}\\
		\dot{p}_\theta&=-\st\,\pphi\left(\ct\,\pphi+2\ppsi\right)+f_\theta\\
		\dot{p}_\phi&=\frac{1}{\st}\left(2\ptheta\,\ppsi+\frac{1}{\st}\left(f_\phi-\ct\,f_\psi\right)\right)\\
		\dot{p}_\psi&=\st\,\ptheta\,\pphi+\frac12 f_\psi-\ct\,\dot{p}_\phi
	\end{align}
	We directly divided the ordinary differential equations (ODEs) of second order into twice as many ODEs of first order. This will be needed for the numerical simulations.
	
	Our notation distinguishes between the time derivative $\dot q$ of a variable, the auxiliary variable $p_q$ (in general not equivalent to the canonical momentum) and later the constants of motion $\vek v_0$ and $\omega_0$. Even though they might have the same numerical values, they represent different physical concepts.
	
	\subsection{Turbulent drag slowing the disk}
	Generally turbulent drag acts proportionally to the fluid density $\rho$, the area $A=\pi r^2$ of the object and its squared velocity. It is always working in the direction opposite to the velocity. In case of translational drag this can be be formulated as
	\begin{align}
	F_x^\text{drag}&=-\frac12 \rho c_x^\text{drag}(\alpha)A|\vek v|\vek v\,,\\
	c_x^\text{drag}(\alpha)&=c_x^0+c_x^\alpha \sin^2\alpha\,.
	\end{align}
	Here $\alpha$ is the angle of attack. It can be calculated via
	\begin{align}
	\alpha\equiv \frac\pi2 -\beta\;,\quad\beta\equiv\arccos\frac{\vd}{|\vek v|}\,.\label{eqn_def_alpha_beta}
	\end{align}
	The coefficient $c_x^0$ gives the resistance at zero angle of attack (see fig.~\ref{fig_disk_sketch_both_sides}) and can be approximated by $c_x^0\approx2rd/A$ where $d$ is the thickness of the disk, though choosing $c_x^0$ slightly larger increases the stability of the simulation significantly. It is not known by how much skin friction increases the coefficient. In contrast the drag coefficient $c_x^\alpha\approx 1.28$ is known from experiment. It acts on the effective area $A\sin\alpha$ with a magnitude again proportional to $\sin\alpha$.
	
	\subsection{Laminar drag damping angular momentum}
	The angular velocities about an axis in the body plane are very low. Due to this it makes sense to consider laminar drag as the main reason for damping of said velocities. The radius of the disk should have cubic influence on this term: two powers from the area, one from the lever. The coefficient is (up to factors $c_\Phi^\text{vis}$ and $c_\Phi^\text{dyn}$ of order one) composed of the dynamic viscosity of air $\eta$ and a dynamic term $|v|r\rho$ reflecting the acceleration of air with a velocity proportional to the distance from the center of rotation, i.e.\ the radius $r$, and accumulating more air, the more is passed during flight, thus yielding an additional factor $|v|$. Put together we receive
	\begin{align}
	F_\theta^\text{lam}&=- \left(c_\Phi^\text{vis}\eta + c_\Phi^\text{dyn}|v|r\rho\right) rA\dtheta\,,\\
	F_\phi^\text{lam}&=- \left(c_\Phi^\text{vis}\eta + c_\Phi^\text{dyn}|v|r\rho\right) rA\st\,\dphi\,.
	\end{align}
	There is a skin friction as well, slowing down the rotation about the symmetry axis. It has a much smaller coefficient but is not negligible on long time scales nevertheless. We write
	\begin{align}
	F_\psi^\text{skin}&=- c_\Phi^\text{skin}\eta rA\left(\dpsi+\ct\dphi\right)\,.
	\end{align}
	
	\subsection{Lifting force}
	Lift is the component of the force on the disk acting perpendicular to the movement. Otherwise the formula is quite similar to the turbulent drag. It has a quadratic dependence in $\vek v$ as well. One finds
	\begin{align}
	F_x^\text{lift}&=\frac12 \rho c_x^\text{lift}(\alpha) A \frac{1}{\sin\beta}\,\vek v\times\left(\vek v\times \vek D\right)\\
	&=\frac12 \rho c_x^\text{lift}(\alpha) A \frac{1}{\sin\beta}\,\left(\left(\vd\right)\vek v-\vek v^2\vek D\right)\\
	&\eqqcolon \tilde{F}^\text{lift}\,\left(\left(\vd\right)\vek v-\vek v^2\vek D\right)\,,\\
	c_x^\text{lift}(\alpha)&=\pi\sin\left(2\alpha\right)
	\end{align}
	where $\alpha$ and $\beta$ as defined in~\eqref{eqn_def_alpha_beta} and $\sin\beta$ is needed for the normalisation of the inner cross-product. The coefficient for thin airfoils has been taken from~\cite{nasa_lift}. As it is known that the lift rapidly drops at about $\alpha\approx20^\circ$, we multiplied the coefficient by a sigmoid
	\begin{align}
	c_x^\text{lift}(\alpha)\mapsto \frac{c_x^\text{lift}(\alpha)}{1+\eto{\left(|\alpha|-\alpha_0\right)/\sigma}}
	\end{align}
	with $\alpha_0=25^\circ$ and $\sigma=5^\circ$. This form is purely qualitative and might differ from reality significantly. However it should not be of high relevance because we expect the flight to proceed mostly with a small angle of attack $\alpha\ll\alpha_0$.
	
	\subsection{Lift-induced torque}
	For a thin airfoil it is known that the aerodynamic center, the point where the lift attacks, lies $25\%$ behind the leading edge~\cite{abbott1959theory}, so the lift introduces an additional torque. For a round disk this means an average distance of $\frac\pi8r$ from the center of mass to the front~\cite{aerodynamik}. Again we have to project the force onto $\vek D$. This projection acts in the $-\vek e$ direction, where
	\begin{equation}
	\vek e = \vek D\times\frac{\vek v\times \vek D}{|\vek v\times \vek D|}=\frac{1}{|\vek v|\sin\beta}\left(\vek v-\left(\vd\right)\vek D\right)
	\end{equation}
	is a unit vector in direction of the leading edge. Putting this together we receive
	\begin{align}
	F_\Phi^\text{lift}&=-\frac\pi8r\left(F_x^\text{lift}\cdot \vek D\right)\vek e\label{eqn_torque_lever}\\
	&=-\frac{\pi}{8}r \tilde{F}^\text{lift}\left(\left(\vd\right)^2-\vek v^2\right)\,\vek e\,,\quad\text{thus}\\
	F_\theta^\text{lift}&=-\frac\pi8r\tilde{F}^\text{lift}\left(\left(\vd\right)^2-\vek v^2\right)\left(\vek e\cdot \vek D_\theta\right)\,,\\
	F_\phi^\text{lift}&=-\frac\pi8r\tilde{F}^\text{lift}\left(\left(\vd\right)^2-\vek v^2\right)\left(\vek e\cdot \vek D_\phi\right)\st\,.
	\end{align}
	
	\subsection{Magnus force}
	The first force that comes in mind when thinking about a rotating object moving through a fluid is the Magnus force~\cite{magnus_effect}. It acts orthogonally to the movement and the angular momentum of the object. In our case the effect is dominated by skin friction and not the movement of air along the rim of the disk, but this should not influence the effect crucially. The resulting force can be written as
	\begin{align}
	F_x^\text{Magnus} &= \rho c_x^\text{Magnus} dA \left(\dpsi+\ct\dphi\right)\, \vek D\times \vek v
	\end{align}
	with a dimensionless coefficient $c_x^\text{Magnus}$ of order one.
	
	\subsection{Convention for the used variables}\label{app:convention}
	
	For the description of a rotating moving disk we choose Cartesian coordinates $\vek x = (x,y,z)$ (vectors are printed in bold) to denote the position and spherical coordinates $(\theta,\,\phi)$ for the orientation of the symmetry axis. The disk rotates about this symmetry axis $\vek D$ by the angle $\psi$. A summary is given in equations~\eqref{eqn_coordinates} and figure~\ref{fig_disk_sketch_both_sides}.
	\begin{align}
		\begin{split}\label{eqn_coordinates}
			\vek x\in \mathbb{R}^3&:\text{position}\\
			\theta\in[0,\,\pi] &: \text{polar angle}\\
			\defined{\phi}{[0,\,2\pi)}{azimuthal angle}\\
			\defined{\psi}{[0,\,2\pi)}{rotation about symmetry axis}\\
			\defined{\vek D}{S^2}{symmetry axis}
		\end{split}
		\\[2ex]
		\begin{split}
			\defined{m}{\mathbb{R}^+}{mass of the disk}\\
			\defined{r}{\mathbb{R}^+}{radius of the disk}\\
			\defined{d}{\mathbb{R}^+}{thickness of the disk}
		\end{split}
		\\[2ex]
		\begin{split}
			\constant{g}{\SI{9.81}{\meter/\second\squared}}{graviation constant}\\
			\constant{\rho}{\SI{1.25}{\kilogram/\meter\cubed}}{density of air}\\
			\constant{\eta}{\SI{18.5}{\micro\pascal\second}}{dynamic viscosity of air}\\
			\defined{\vek T}{\mathbb{R}^3}{symmetry axis preferred by external torque}\\
			\defined{c_q^i}{\mathbb{R}^+}{proportionality coefficient in the force $F_q^i$}
		\end{split}
		\\[2ex]
		\begin{split}
			\defined{\alpha}{[0,\pi/2]}{angle of attack}\\
			\defined{\beta}{[0,\pi]}{angle between velocity and symmetry axis}\\
			\defined{\vek e}{S^2}{direction of the leading edge}
		\end{split}
	\end{align}

%% file: sections/numerical_solution.tex
\section{Numerical simulation}\label{sec:numerical_simulation}
	In order to test the considerations from the previous section we solve the system of differential equations numerically. It turns out that this is much more challenging than expected. The results presented here do not describe the flight of a disk quantitatively correctly for several reasons. First of all this is not our goal, at least for now. We are only interested in a numerical verification of our qualitative arguments. In addition we do not know many of the coefficients without experiment, the forces we use are only approximations and it is not clear if we left out relevant terms.
	\subsection{Algorithms}
	A huge issue for numerical stability is the region where $\theta\approx 0$. The two points where $\theta=0$ exactly are null sets. Therefore they are protected by a centrifugal barrier and are never reached. This leads to divergent forces $\sim\sin^{-3}\theta$. The stability can be improved significantly via coordinate transformations into a system where the polar angle is measured from the axis $x$, $y$, or $z$ with the smallest value $d_i$, i.e.\@ the largest corresponding $\theta$. For algorithmic details see appendix~\ref{app:coord_trafo}. Fortunately the coordinate transformation leaves the formulae apart from equation~\eqref{eqn_def_d} forminvariant. Only the dependence of $\vek D$ on the angles changes.

	Depending on the choice of the coefficients the differential equations sometimes turn out to be stiff~\cite{wanner_ode_stiff}. For this reason we implemented two different numeric integrators with adaptive step size. In all physically interesting cases the Dormand-Prince 5(4) algorithm (DOPRI54)~\cite{dormand_prince,wanner_ode_nonstiff} gives good results. It is a non-stiff integrator of order 5 that uses an embedded 4th order integrator for step size correction. In addition we implemented the Rosenbrock-Wanner 2(3) algorithm (ROW23)~\cite{matlab_ode} of order two with third order step size correction. It is suitable for stiff ODEs being A- and L-stable. We cross checked the correctness of the usually faster DOPRI54 with the stable ROW23.

	\subsection{Results}
	If not mentioned otherwise explicitly, the following results all use the parameters from equations~\eqref{eqn_constant_coeffs} and~\eqref{eqn_initial_params} in \cref{app:params}. We tried to keep the values as near to the ones of a beer mat as possible, but we had to guess several values. This is especially critical for $c_x^0$, $c_x^\text{Magnus}$ and $c_\Phi^\text{dyn}$ where we did not find good literature values, but a strong dependency of the flight behaviour.
	
	We did not attempt any fine tuning of these parameters, on the contrary, we restricted ourself to integer or inverse integer values, because we are mostly interested in the qualitatively correct behaviour. It is an important stability test that minor changes in the parameters do not significantly change the picture as a whole. Otherwise our predictions might not be valid on a mountain with thinner air, on a hot or humid day, or for a slightly thicker disk. For these reasons our numerical simulations do not match the experimental data exactly (see figs.~\ref{fig_x_t_trajectory} and~\ref{fig_y_t_trajectory}). Nevertheless they provide a reasonably good approximation capturing all the important features and providing a proof that we identified the relevant forces.

\begin{figure}[ht]
		\centering
		\input{figures/lift_magnus_phi_th}
		\caption{Time evolution of a disk simulated with the relevant force terms. The angles are normalised in such a way that the zero-line marks backspin.}\label{fig_lift_magnus}
	\end{figure}
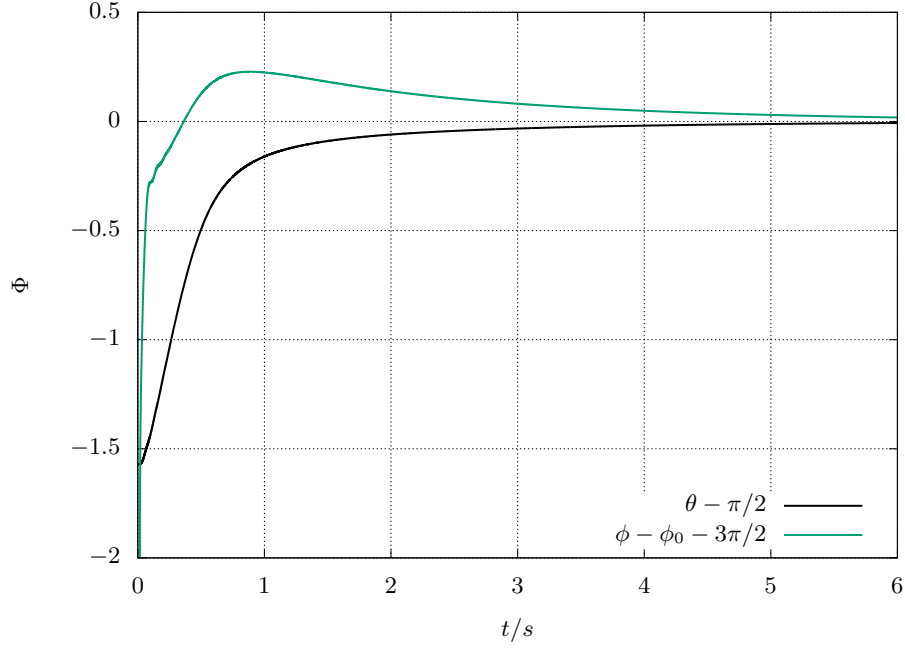

	Figure~\ref{fig_lift_magnus} shows the trajectory of a disk with all the relevant forces given in section~\ref{sec:forces} included. We chose to shift the angles by $\pi/2$ and $3\pi/2$ respectively, so that the zero-line marks backspin. In addition we normalised the azimuthal angle $\phi$ to the flight direction $\phi_0$. One clearly sees that the disk converges to backspin orientation. The time scale of about half a second on which this happens is realistic. Once backspin has been reached, the disk remains in this orientation for any simulated duration. It is also interesting to observe that $\phi$ reaches the vicinity of its target value much earlier than $\theta$. While $\theta$ is smoothly and monotonously increasing, $\phi$ reaches an orientation of minimised air resistance practically immediately. Afterwards $\phi$ overshoots its final value and continues a with a smooth convergence. This overshooting is caused by the positive angle of attack and is crucial for the convergence towards backspin.

\begin{figure}[ht]
		\centering
		\input{figures/only_magnus_phi_th}
		\caption{Time evolution of a disk neglecting the lift-induced torque ($F_\Phi^\text{lift}=0$). The angles are normalised in such a way that the zero-line marks backspin. The discontinuity comes from projecting $\phi-\phi_0$ onto the interval $[0,2\pi)$.}\label{fig_magnus}
	\end{figure}
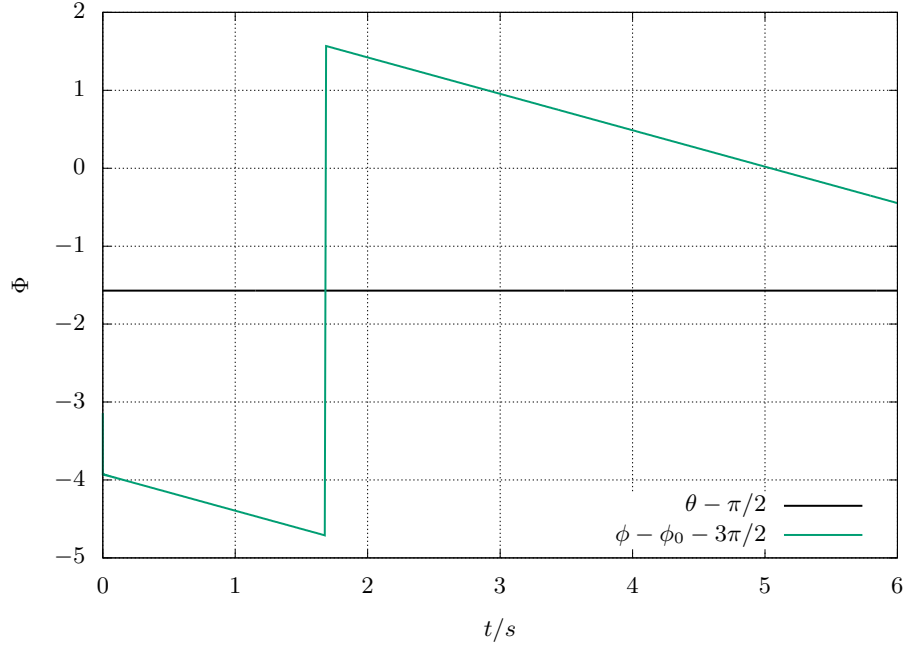

	Next we show that $F_\Phi^\text{lift}$ is indeed the force responsible for always converging to backspin. In figure~\ref{fig_magnus} we visualise the time evolution in absence of this force. In this case the polar angle $\theta$ does not change and the azimuthal angle $\phi$ does not carry any meaning at $\theta=0$. In other words, there is no torque and the orientation remains as it is without ever reaching backspin.
	As there is no difference in the initial conditions leading to the figures~\ref{fig_lift_magnus} and~\ref{fig_magnus} respectively other than $c_\Phi^\text{lever}$, we can conclude that it is this effect that leads to backspin as the only preferred orientation.

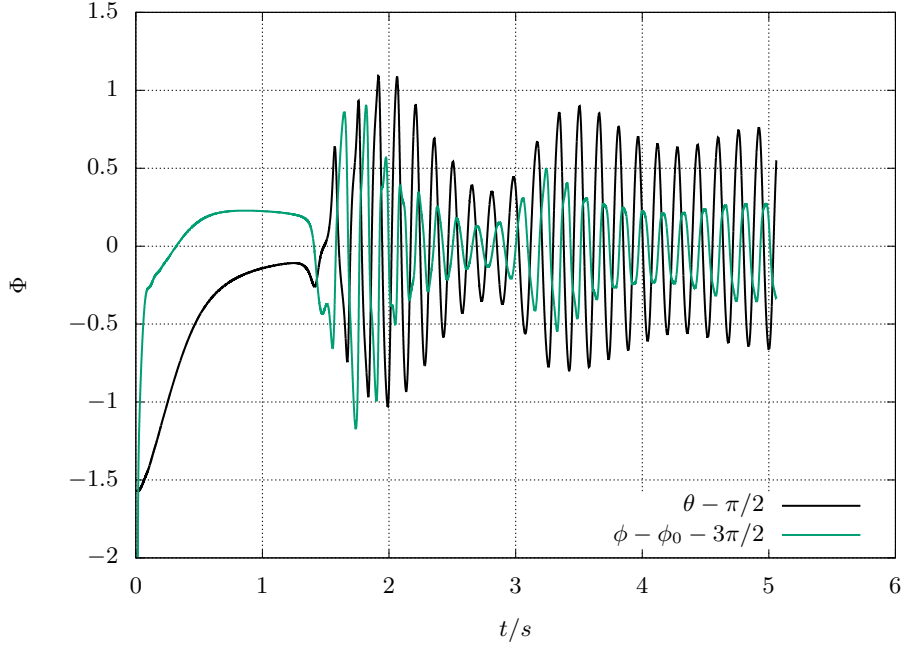
\begin{figure}[ht]
		\centering
		\input{figures/only_lift_phi_th}
		\caption{Time evolution of a disk neglecting the Magnus effect and the drag of a perfectly aligned disk (i.e.\ $c_x^\text{Magnus}=0$ and $c_x^0=0$). The angles are normalised in such a way that the zero-line marks backspin.}\label{fig_lift}
	\end{figure}

	Let us conclude with an example of coefficients chosen in such a way that the orientation of the disk does not converge at all. For this we show the time evolution without the Magnus effect and the drag at zero angle of attack in figure~\ref{fig_lift}. The system does not converge because damping and gravitational acceleration just balance each other. Instead it oscillates about the backspin orientation. Again we find that $F_\Phi^\text{lift}$ leads to a preference of this orientation, thus we do not need the Magnus effect in principle to explain the qualitative trend. From this and other simulations we find, however, that it is crucial for the stability.

	\subsection{Details on the coordinate transformation}\label{app:coord_trafo}
	Here we denote the spherical coordinates with reference to $x_k$ by $(\theta_k,\,\phi_k)$. I.e.\@ in the usual case we use $z\equiv x_3$ and have $(\theta,\,\phi)\equiv(\theta_3,\,\phi_3)$. Then equation~\eqref{eqn_def_d} generalises to
	\begin{alignat}{6}
		\vek D&=\matr{c}{\cp_3\st_3\\\sp_3\st_3\\\ct_3}&=&\matr{c}{\ct_1\\\cp_1\st_1\\\sp_1\st_1}&=&\matr{c}{\sp_2\st_2\\\ct_2\\\cp_2\st_2}\,,\label{eqn_def_d_gen}\\
		\vek D_\theta&=\matr{c}{\cp_3\ct_3\\\sp_3\ct_3\\-\st_3}&=&\matr{c}{-\st_1\\\cp_1 \ct_1\\\ct_1 \sp_1}&=&\matr{c}{\ct_2 \sp_2\\-\st_2\\\cp_2 \ct_2}\,,\\
		\vek D_\phi&=\matr{c}{-\sp_3\\\cp_3\\0}&=&\matr{c}{0\\-\sp_1\\\cp_1}&=&\matr{c}{\cp_2\\0\\-\sp_2}\,.
	\end{alignat}
	Then a coordinate transformation from the initial axis $x_i$ to the final one $x_f$ in given in algorithm~\ref{alg_coord_trafo}. The formulae for $\theta$ and $\phi$ are simply the geometrical definitions. The derivatives of these formulae directly lead to the equations for $\ptheta=\dtheta$ and $\pphi=\dphi$.
	\begin{algorithm}
		\SetKwInOut{Input}{input}\SetKwInOut{Output}{output}
		\Input{$i$, $f$, $(\theta_i,\,\phi_i)$, $(\dtheta_i,\,\dphi_i,\,\dpsi_i)$}
		\Output{$(\theta_f,\,\phi_f)$, $(\dtheta_f,\,\dphi_f,\,\dpsi_f)$}
		\BlankLine
		Calculate $\vek D$ via eq.~\eqref{eqn_def_d_gen}\;
		$L_3=\ct_i\dphi_i+\dpsi$ \tcp*{conserved angular momentum}
		$\theta_f=\arccos D_f$\;
		$\phi_f=\atan\left(D_{f+2},\,D_{f+1}\right)$ \tcp*{the shift is a permutation}
		\If{$(i,f,r)$ even permutation}{
			$\dtheta_f = \left(\sp_i\st_i\dphi_i-\cp_i\ct_i\dtheta_i\right)/\sqrt{1-\cos^2\phi_i\sin^2\theta_i}$\;
			$\dphi_f = -\left(\sp_i\dtheta_i+\cp_i\ct_i\st_i\dphi_i\right)/\left(\cos^2\theta_i+\sin^2\phi_i\sin^2\theta_i\right)$\;
		}\Else{
			$\dtheta_f = -\left(\ct_i\sp_i\dtheta_i+\cp_i\st_i\dphi_i\right)/\sqrt{1-\sin^2\phi_i\sin^2\theta_i}$\;
			$\dphi_f = \left(\cp_i\dtheta_i-\ct_i\sp_i\st_i\dphi_i\right)/\left(\cos^2\theta_i+\cos^2\phi_i\sin^2\theta_i\right)$\;
		}
		$\dpsi_f=L_3-\ct_f\dphi_f$\;
		\caption{Coordinate transformation from $x_i$ to $x_f$ (identifying $(x_1,x_2,x_3)\equiv(x,y,z)$). We call the remaining axis $x_r$. The vector $\vek D$ does not change.}\label{alg_coord_trafo}
	\end{algorithm}
	
	\subsection{Parameters used in the numerical simulation}\label{app:params}
	The coefficients listed in equation~\eqref{eqn_constant_coeffs} reproduce the values of a beer mat as accurately as possible while setting the maximum number of coefficients to zero and without fine tuning. The non-zero coefficients are
	\begin{equation}
		\begin{split}
			m&=\SI{5.9}{\gram}\,,\\
			r&=\SI{5.3}{\centi\meter}\,,\\
			d&=\SI{1.7}{\milli\meter}\,,\\
			c^0_x&=\num{0.2}\,,\\
			c_{\vek x}^\alpha&=\num{1.28}\,,\\
			c_{\vek x}^\text{Magnus}&=\num{3}\,,\\
			c_\Phi^\text{vis}&=\num{1}\,,\\
			c_\Phi^\text{dyn}&=\num{0.5}\,.
		\end{split}\label{eqn_constant_coeffs}
	\end{equation}
	We deliberately chose to trade some accuracy in the description for better qualitative understanding by removing all the terms that are not needed to describe the investigated effect.
	
	The relevant starting parameters have been set, again reproducing the experimental setup as realistically as possible, to
	\begin{align}
		\begin{split}
			\theta &= 0\,,\\
			\phi &= 0\,,\\
			x &= 0\,,\\
			y &= 0\,,\\
			z &= \SI{1}{\meter}\,,\\
			\dot x &= \SI{2.62}{\meter/\second}\,,\\
			\dot y &= \SI{0}{\meter/\second}\,,\\
			\dot z &= \SI{0.05}{\meter/\second}\,,\\
			\dpsi &= \SI{49}{\radian/\second}\,.
		\end{split}\label{eqn_initial_params}
	\end{align}
	We checked that different initial angles do not influence the long term behaviour by repeating the simulation several times with $\vek D$ sampled randomly from a uniform distribution on the unit sphere.

%% file: figures/lift_magnus_phi_th.tex
\begingroup
  \inputencoding{latin1}%
  \makeatletter
  \providecommand\color[2][]{%
    \GenericError{(gnuplot) \space\space\space\@spaces}{%
      Package color not loaded in conjunction with
      terminal option `colourtext'%
    }{See the gnuplot documentation for explanation.%
    }{Either use 'blacktext' in gnuplot or load the package
      color.sty in LaTeX.}%
    \renewcommand\color[2][]{}%
  }%
  \providecommand\includegraphics[2][]{%
    \GenericError{(gnuplot) \space\space\space\@spaces}{%
      Package graphicx or graphics not loaded%
    }{See the gnuplot documentation for explanation.%
    }{The gnuplot epslatex terminal needs graphicx.sty or graphics.sty.}%
    \renewcommand\includegraphics[2][]{}%
  }%
  \providecommand\rotatebox[2]{#2}%
  \@ifundefined{ifGPcolor}{%
    \newif\ifGPcolor
    \GPcolortrue
  }{}%
  \@ifundefined{ifGPblacktext}{%
    \newif\ifGPblacktext
    \GPblacktexttrue
  }{}%
  \let\gplgaddtomacro\g@addto@macro
  \gdef\gplbacktext{}%
  \gdef\gplfronttext{}%
  \makeatother
  \ifGPblacktext
    \def\colorrgb#1{}%
    \def\colorgray#1{}%
  \else
    \ifGPcolor
      \def\colorrgb#1{\color[rgb]{#1}}%
      \def\colorgray#1{\color[gray]{#1}}%
      \expandafter\def\csname LTw\endcsname{\color{white}}%
      \expandafter\def\csname LTb\endcsname{\color{black}}%
      \expandafter\def\csname LTa\endcsname{\color{black}}%
      \expandafter\def\csname LT0\endcsname{\color[rgb]{1,0,0}}%
      \expandafter\def\csname LT1\endcsname{\color[rgb]{0,1,0}}%
      \expandafter\def\csname LT2\endcsname{\color[rgb]{0,0,1}}%
      \expandafter\def\csname LT3\endcsname{\color[rgb]{1,0,1}}%
      \expandafter\def\csname LT4\endcsname{\color[rgb]{0,1,1}}%
      \expandafter\def\csname LT5\endcsname{\color[rgb]{1,1,0}}%
      \expandafter\def\csname LT6\endcsname{\color[rgb]{0,0,0}}%
      \expandafter\def\csname LT7\endcsname{\color[rgb]{1,0.3,0}}%
      \expandafter\def\csname LT8\endcsname{\color[rgb]{0.5,0.5,0.5}}%
    \else
      \def\colorrgb#1{\color{black}}%
      \def\colorgray#1{\color[gray]{#1}}%
      \expandafter\def\csname LTw\endcsname{\color{white}}%
      \expandafter\def\csname LTb\endcsname{\color{black}}%
      \expandafter\def\csname LTa\endcsname{\color{black}}%
      \expandafter\def\csname LT0\endcsname{\color{black}}%
      \expandafter\def\csname LT1\endcsname{\color{black}}%
      \expandafter\def\csname LT2\endcsname{\color{black}}%
      \expandafter\def\csname LT3\endcsname{\color{black}}%
      \expandafter\def\csname LT4\endcsname{\color{black}}%
      \expandafter\def\csname LT5\endcsname{\color{black}}%
      \expandafter\def\csname LT6\endcsname{\color{black}}%
      \expandafter\def\csname LT7\endcsname{\color{black}}%
      \expandafter\def\csname LT8\endcsname{\color{black}}%
    \fi
  \fi
    \setlength{\unitlength}{0.0500bp}%
    \ifx\gptboxheight\undefined%
      \newlength{\gptboxheight}%
      \newlength{\gptboxwidth}%
      \newsavebox{\gptboxtext}%
    \fi%
    \setlength{\fboxrule}{0.5pt}%
    \setlength{\fboxsep}{1pt}%
\begin{picture}(7200.00,5040.00)%
    \gplgaddtomacro\gplbacktext{%
      \csname LTb\endcsname
      \put(946,704){\makebox(0,0)[r]{\strut{}$-2$}}%
      \csname LTb\endcsname
      \put(946,1527){\makebox(0,0)[r]{\strut{}$-1.5$}}%
      \csname LTb\endcsname
      \put(946,2350){\makebox(0,0)[r]{\strut{}$-1$}}%
      \csname LTb\endcsname
      \put(946,3173){\makebox(0,0)[r]{\strut{}$-0.5$}}%
      \csname LTb\endcsname
      \put(946,3996){\makebox(0,0)[r]{\strut{}$0$}}%
      \csname LTb\endcsname
      \put(946,4819){\makebox(0,0)[r]{\strut{}$0.5$}}%
      \csname LTb\endcsname
      \put(1078,484){\makebox(0,0){\strut{}$0$}}%
      \csname LTb\endcsname
      \put(2032,484){\makebox(0,0){\strut{}$1$}}%
      \csname LTb\endcsname
      \put(2986,484){\makebox(0,0){\strut{}$2$}}%
      \csname LTb\endcsname
      \put(3941,484){\makebox(0,0){\strut{}$3$}}%
      \csname LTb\endcsname
      \put(4895,484){\makebox(0,0){\strut{}$4$}}%
      \csname LTb\endcsname
      \put(5849,484){\makebox(0,0){\strut{}$5$}}%
      \csname LTb\endcsname
      \put(6803,484){\makebox(0,0){\strut{}$6$}}%
    }%
    \gplgaddtomacro\gplfronttext{%
      \csname LTb\endcsname
      \put(209,2761){\rotatebox{-270}{\makebox(0,0){\strut{}$\Phi$}}}%
      \put(3940,154){\makebox(0,0){\strut{}$t/s$}}%
      \csname LTb\endcsname
      \put(5816,1097){\makebox(0,0)[r]{\strut{}$\theta-\pi/2$}}%
      \csname LTb\endcsname
      \put(5816,877){\makebox(0,0)[r]{\strut{}$\phi-\phi_0-3\pi/2$}}%
    }%
    \gplbacktext
    \put(0,0){\includegraphics{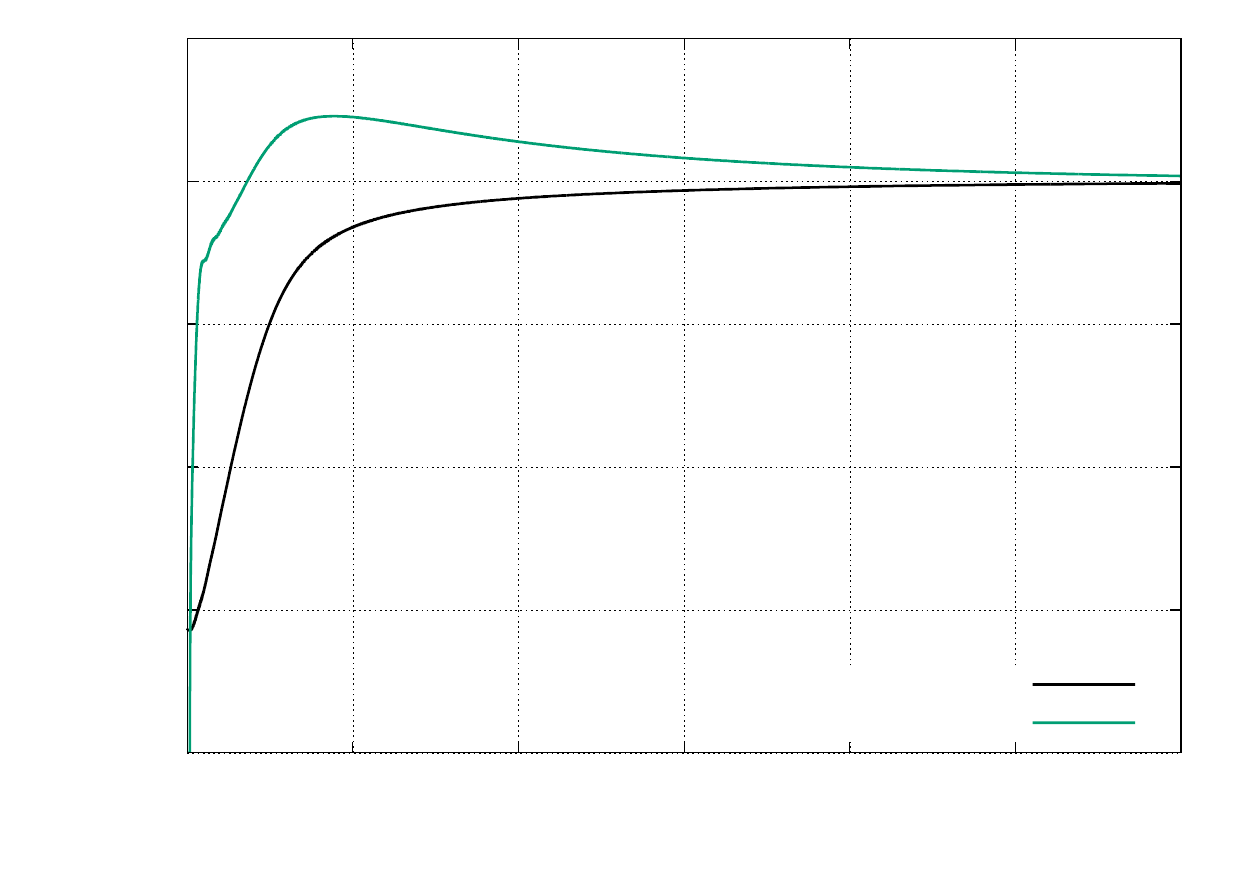}}%
    \gplfronttext
  \end{picture}%
\endgroup

%% file: figures/only_magnus_phi_th.tex
\begingroup
  \inputencoding{latin1}%
  \makeatletter
  \providecommand\color[2][]{%
    \GenericError{(gnuplot) \space\space\space\@spaces}{%
      Package color not loaded in conjunction with
      terminal option `colourtext'%
    }{See the gnuplot documentation for explanation.%
    }{Either use 'blacktext' in gnuplot or load the package
      color.sty in LaTeX.}%
    \renewcommand\color[2][]{}%
  }%
  \providecommand\includegraphics[2][]{%
    \GenericError{(gnuplot) \space\space\space\@spaces}{%
      Package graphicx or graphics not loaded%
    }{See the gnuplot documentation for explanation.%
    }{The gnuplot epslatex terminal needs graphicx.sty or graphics.sty.}%
    \renewcommand\includegraphics[2][]{}%
  }%
  \providecommand\rotatebox[2]{#2}%
  \@ifundefined{ifGPcolor}{%
    \newif\ifGPcolor
    \GPcolortrue
  }{}%
  \@ifundefined{ifGPblacktext}{%
    \newif\ifGPblacktext
    \GPblacktexttrue
  }{}%
  \let\gplgaddtomacro\g@addto@macro
  \gdef\gplbacktext{}%
  \gdef\gplfronttext{}%
  \makeatother
  \ifGPblacktext
    \def\colorrgb#1{}%
    \def\colorgray#1{}%
  \else
    \ifGPcolor
      \def\colorrgb#1{\color[rgb]{#1}}%
      \def\colorgray#1{\color[gray]{#1}}%
      \expandafter\def\csname LTw\endcsname{\color{white}}%
      \expandafter\def\csname LTb\endcsname{\color{black}}%
      \expandafter\def\csname LTa\endcsname{\color{black}}%
      \expandafter\def\csname LT0\endcsname{\color[rgb]{1,0,0}}%
      \expandafter\def\csname LT1\endcsname{\color[rgb]{0,1,0}}%
      \expandafter\def\csname LT2\endcsname{\color[rgb]{0,0,1}}%
      \expandafter\def\csname LT3\endcsname{\color[rgb]{1,0,1}}%
      \expandafter\def\csname LT4\endcsname{\color[rgb]{0,1,1}}%
      \expandafter\def\csname LT5\endcsname{\color[rgb]{1,1,0}}%
      \expandafter\def\csname LT6\endcsname{\color[rgb]{0,0,0}}%
      \expandafter\def\csname LT7\endcsname{\color[rgb]{1,0.3,0}}%
      \expandafter\def\csname LT8\endcsname{\color[rgb]{0.5,0.5,0.5}}%
    \else
      \def\colorrgb#1{\color{black}}%
      \def\colorgray#1{\color[gray]{#1}}%
      \expandafter\def\csname LTw\endcsname{\color{white}}%
      \expandafter\def\csname LTb\endcsname{\color{black}}%
      \expandafter\def\csname LTa\endcsname{\color{black}}%
      \expandafter\def\csname LT0\endcsname{\color{black}}%
      \expandafter\def\csname LT1\endcsname{\color{black}}%
      \expandafter\def\csname LT2\endcsname{\color{black}}%
      \expandafter\def\csname LT3\endcsname{\color{black}}%
      \expandafter\def\csname LT4\endcsname{\color{black}}%
      \expandafter\def\csname LT5\endcsname{\color{black}}%
      \expandafter\def\csname LT6\endcsname{\color{black}}%
      \expandafter\def\csname LT7\endcsname{\color{black}}%
      \expandafter\def\csname LT8\endcsname{\color{black}}%
    \fi
  \fi
    \setlength{\unitlength}{0.0500bp}%
    \ifx\gptboxheight\undefined%
      \newlength{\gptboxheight}%
      \newlength{\gptboxwidth}%
      \newsavebox{\gptboxtext}%
    \fi%
    \setlength{\fboxrule}{0.5pt}%
    \setlength{\fboxsep}{1pt}%
\begin{picture}(7200.00,5040.00)%
    \gplgaddtomacro\gplbacktext{%
      \csname LTb\endcsname
      \put(682,704){\makebox(0,0)[r]{\strut{}$-5$}}%
      \csname LTb\endcsname
      \put(682,1292){\makebox(0,0)[r]{\strut{}$-4$}}%
      \csname LTb\endcsname
      \put(682,1880){\makebox(0,0)[r]{\strut{}$-3$}}%
      \csname LTb\endcsname
      \put(682,2468){\makebox(0,0)[r]{\strut{}$-2$}}%
      \csname LTb\endcsname
      \put(682,3055){\makebox(0,0)[r]{\strut{}$-1$}}%
      \csname LTb\endcsname
      \put(682,3643){\makebox(0,0)[r]{\strut{}$0$}}%
      \csname LTb\endcsname
      \put(682,4231){\makebox(0,0)[r]{\strut{}$1$}}%
      \csname LTb\endcsname
      \put(682,4819){\makebox(0,0)[r]{\strut{}$2$}}%
      \csname LTb\endcsname
      \put(814,484){\makebox(0,0){\strut{}$0$}}%
      \csname LTb\endcsname
      \put(1812,484){\makebox(0,0){\strut{}$1$}}%
      \csname LTb\endcsname
      \put(2810,484){\makebox(0,0){\strut{}$2$}}%
      \csname LTb\endcsname
      \put(3809,484){\makebox(0,0){\strut{}$3$}}%
      \csname LTb\endcsname
      \put(4807,484){\makebox(0,0){\strut{}$4$}}%
      \csname LTb\endcsname
      \put(5805,484){\makebox(0,0){\strut{}$5$}}%
      \csname LTb\endcsname
      \put(6803,484){\makebox(0,0){\strut{}$6$}}%
    }%
    \gplgaddtomacro\gplfronttext{%
      \csname LTb\endcsname
      \put(209,2761){\rotatebox{-270}{\makebox(0,0){\strut{}$\Phi$}}}%
      \put(3808,154){\makebox(0,0){\strut{}$t/s$}}%
      \csname LTb\endcsname
      \put(5816,1097){\makebox(0,0)[r]{\strut{}$\theta-\pi/2$}}%
      \csname LTb\endcsname
      \put(5816,877){\makebox(0,0)[r]{\strut{}$\phi-\phi_0-3\pi/2$}}%
    }%
    \gplbacktext
    \put(0,0){\includegraphics{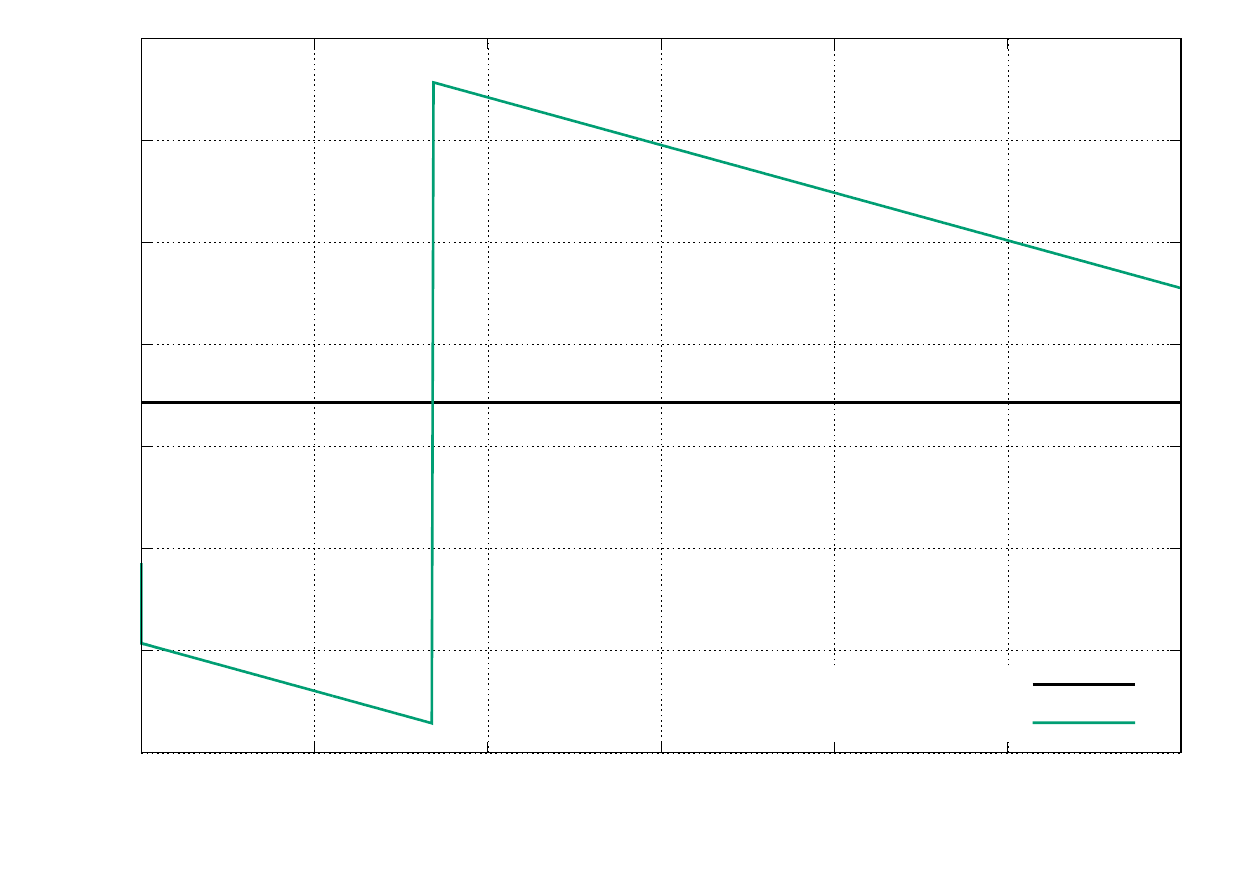}}%
    \gplfronttext
  \end{picture}%
\endgroup

%% file: figures/only_lift_phi_th.tex
\begingroup
  \inputencoding{latin1}%
  \makeatletter
  \providecommand\color[2][]{%
    \GenericError{(gnuplot) \space\space\space\@spaces}{%
      Package color not loaded in conjunction with
      terminal option `colourtext'%
    }{See the gnuplot documentation for explanation.%
    }{Either use 'blacktext' in gnuplot or load the package
      color.sty in LaTeX.}%
    \renewcommand\color[2][]{}%
  }%
  \providecommand\includegraphics[2][]{%
    \GenericError{(gnuplot) \space\space\space\@spaces}{%
      Package graphicx or graphics not loaded%
    }{See the gnuplot documentation for explanation.%
    }{The gnuplot epslatex terminal needs graphicx.sty or graphics.sty.}%
    \renewcommand\includegraphics[2][]{}%
  }%
  \providecommand\rotatebox[2]{#2}%
  \@ifundefined{ifGPcolor}{%
    \newif\ifGPcolor
    \GPcolortrue
  }{}%
  \@ifundefined{ifGPblacktext}{%
    \newif\ifGPblacktext
    \GPblacktexttrue
  }{}%
  \let\gplgaddtomacro\g@addto@macro
  \gdef\gplbacktext{}%
  \gdef\gplfronttext{}%
  \makeatother
  \ifGPblacktext
    \def\colorrgb#1{}%
    \def\colorgray#1{}%
  \else
    \ifGPcolor
      \def\colorrgb#1{\color[rgb]{#1}}%
      \def\colorgray#1{\color[gray]{#1}}%
      \expandafter\def\csname LTw\endcsname{\color{white}}%
      \expandafter\def\csname LTb\endcsname{\color{black}}%
      \expandafter\def\csname LTa\endcsname{\color{black}}%
      \expandafter\def\csname LT0\endcsname{\color[rgb]{1,0,0}}%
      \expandafter\def\csname LT1\endcsname{\color[rgb]{0,1,0}}%
      \expandafter\def\csname LT2\endcsname{\color[rgb]{0,0,1}}%
      \expandafter\def\csname LT3\endcsname{\color[rgb]{1,0,1}}%
      \expandafter\def\csname LT4\endcsname{\color[rgb]{0,1,1}}%
      \expandafter\def\csname LT5\endcsname{\color[rgb]{1,1,0}}%
      \expandafter\def\csname LT6\endcsname{\color[rgb]{0,0,0}}%
      \expandafter\def\csname LT7\endcsname{\color[rgb]{1,0.3,0}}%
      \expandafter\def\csname LT8\endcsname{\color[rgb]{0.5,0.5,0.5}}%
    \else
      \def\colorrgb#1{\color{black}}%
      \def\colorgray#1{\color[gray]{#1}}%
      \expandafter\def\csname LTw\endcsname{\color{white}}%
      \expandafter\def\csname LTb\endcsname{\color{black}}%
      \expandafter\def\csname LTa\endcsname{\color{black}}%
      \expandafter\def\csname LT0\endcsname{\color{black}}%
      \expandafter\def\csname LT1\endcsname{\color{black}}%
      \expandafter\def\csname LT2\endcsname{\color{black}}%
      \expandafter\def\csname LT3\endcsname{\color{black}}%
      \expandafter\def\csname LT4\endcsname{\color{black}}%
      \expandafter\def\csname LT5\endcsname{\color{black}}%
      \expandafter\def\csname LT6\endcsname{\color{black}}%
      \expandafter\def\csname LT7\endcsname{\color{black}}%
      \expandafter\def\csname LT8\endcsname{\color{black}}%
    \fi
  \fi
    \setlength{\unitlength}{0.0500bp}%
    \ifx\gptboxheight\undefined%
      \newlength{\gptboxheight}%
      \newlength{\gptboxwidth}%
      \newsavebox{\gptboxtext}%
    \fi%
    \setlength{\fboxrule}{0.5pt}%
    \setlength{\fboxsep}{1pt}%
\begin{picture}(7200.00,5040.00)%
    \gplgaddtomacro\gplbacktext{%
      \csname LTb\endcsname
      \put(946,704){\makebox(0,0)[r]{\strut{}$-2$}}%
      \csname LTb\endcsname
      \put(946,1292){\makebox(0,0)[r]{\strut{}$-1.5$}}%
      \csname LTb\endcsname
      \put(946,1880){\makebox(0,0)[r]{\strut{}$-1$}}%
      \csname LTb\endcsname
      \put(946,2468){\makebox(0,0)[r]{\strut{}$-0.5$}}%
      \csname LTb\endcsname
      \put(946,3055){\makebox(0,0)[r]{\strut{}$0$}}%
      \csname LTb\endcsname
      \put(946,3643){\makebox(0,0)[r]{\strut{}$0.5$}}%
      \csname LTb\endcsname
      \put(946,4231){\makebox(0,0)[r]{\strut{}$1$}}%
      \csname LTb\endcsname
      \put(946,4819){\makebox(0,0)[r]{\strut{}$1.5$}}%
      \csname LTb\endcsname
      \put(1078,484){\makebox(0,0){\strut{}$0$}}%
      \csname LTb\endcsname
      \put(2032,484){\makebox(0,0){\strut{}$1$}}%
      \csname LTb\endcsname
      \put(2986,484){\makebox(0,0){\strut{}$2$}}%
      \csname LTb\endcsname
      \put(3941,484){\makebox(0,0){\strut{}$3$}}%
      \csname LTb\endcsname
      \put(4895,484){\makebox(0,0){\strut{}$4$}}%
      \csname LTb\endcsname
      \put(5849,484){\makebox(0,0){\strut{}$5$}}%
      \csname LTb\endcsname
      \put(6803,484){\makebox(0,0){\strut{}$6$}}%
    }%
    \gplgaddtomacro\gplfronttext{%
      \csname LTb\endcsname
      \put(209,2761){\rotatebox{-270}{\makebox(0,0){\strut{}$\Phi$}}}%
      \put(3940,154){\makebox(0,0){\strut{}$t/s$}}%
      \csname LTb\endcsname
      \put(5816,1097){\makebox(0,0)[r]{\strut{}$\theta-\pi/2$}}%
      \csname LTb\endcsname
      \put(5816,877){\makebox(0,0)[r]{\strut{}$\phi-\phi_0-3\pi/2$}}%
    }%
    \gplbacktext
    \put(0,0){\includegraphics{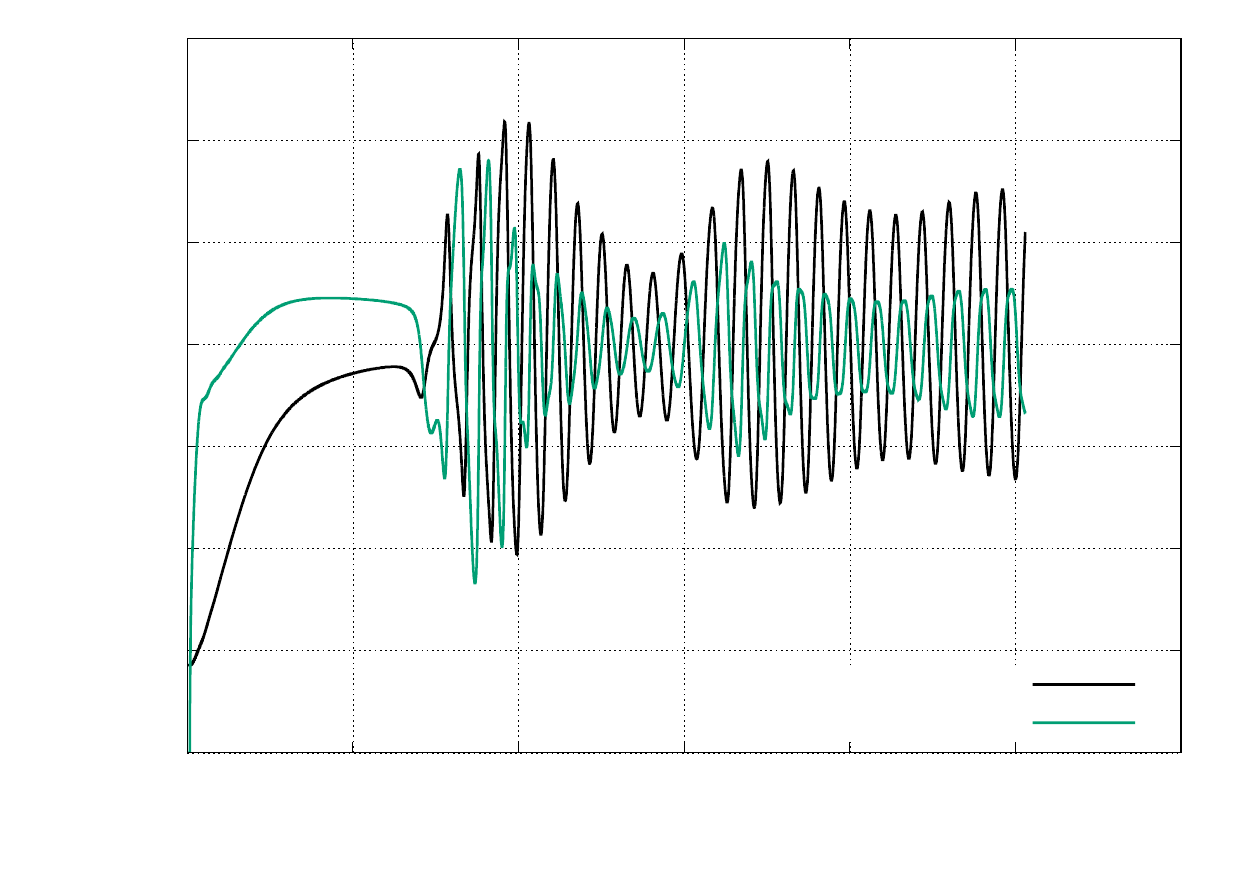}}%
    \gplfronttext
  \end{picture}%
\endgroup

%% file: sections/more_verification.tex
\section{The $x$- and $z$-components in the effective theory}\label{sec_x_and_z_eff_theory}
	In the main article we solve the differential equation obtained from the effective theory only for the $y$-component of the trajectory (see eq.~\eqref{eqn_eff_sol_x2}). This coordinate shows the most remarkable behaviour because it is dominated by two very distinct regimes and the time scale $\tau$ of the rotation features prominently in the curve. Nevertheless the $x$- and $z$-components are of interest as well and we are going to investigate them here closely.
	
 	The differential equation
 	\begin{align}
 	\ddot x &= -c^\perp_x\frac Am \dot x\\
 	\Rightarrow x(t) &= v_0\tau_x\left(1-\eto{-t/\tau_x}\right)\,,\label{eqn_eff_sol_x1}\\
 	\tau_x &\coloneqq \frac{m}{c^\perp_x A}
 	\end{align}
 	is easily solvable for $x$ by integration.
 	
	Last but not least we investigate the vertical component $z$ following the differential equation
	\begin{align}
	\ddot z &= -g\left(1-\varepsilon\cos^2\vartheta\right)-c^\text{damp}_x(\vartheta)\frac Am \dot z\,.
 	\end{align}
 	Here we introduced a dependence of the damping coefficient $c^\text{damp}_x(\vartheta)$ on the angle $\vartheta$. We do not make any predictions about the nature of this dependence other than the damping being larger parallel to $\vek D$ than orthogonal to $\vek D$, writing $c^\text{damp}_x(0) \eqqcolon c^{\parallel}_x>c^{\perp}_x \coloneqq c^\text{damp}_x(\pm\pi/2)$.
 	
 	For convenience let us define the time scale $\tau_z\coloneqq \frac{m}{c^{\parallel}_x A}$ and the coefficient $c^\text{rel}_x\coloneqq c^{\parallel}_x/c^{\perp}_x$. As always we start with side spin. Similarly to the considerations of the $y$-component we can split the dynamics into two stages that have to be connected smoothly. In the first one ($t\ll \tau$) we use $\cos^2\vartheta\approx 1$ and $c^\text{damp}_x(\vartheta)\approx c^{\parallel}_x$, in the second one ($t\gg \tau$) the other extreme $\cos^2\vartheta\approx 0$ and $c^\text{damp}_x(\vartheta)\approx c^{\perp}_x$ applies. In both stages the differential equation is easily solvable and we use the same transition function as in equation~\eqref{eqn_eff_sol_x2} to obtain
 	\begin{align}
 	z(t) &\approx h - g(1-\varepsilon)\tau_z\left(t-\tau_z\left(1-\eto{-t/\tau_z}\right)\right)
 	- g\left(c^\text{rel}_x-1+\varepsilon\right)\tau_zt_i\log\left(1+\eto{\left(t-\tau\right)/t_i}\right)\label{eqn_eff_sol_x3}\\
 	&\approx \begin{cases}
 	h - \frac 12g(1-\varepsilon)t^2 & t<\tau_z\\
 	h - g(1-\varepsilon)\tau_z\left(t-\tau_z\right) & \tau_z < t < \tau\\
 	h - gc^\text{rel}_x\tau_z\left(t-\left(\tau-\frac{1-\varepsilon}{c^\text{rel}_x}\left(\tau-\tau_z\right)\right)\right) & \tau < t
 	\end{cases}
 	\end{align}
 	where we assumed $\tau_z<\tau$ which we find to hold for realistic experiments.

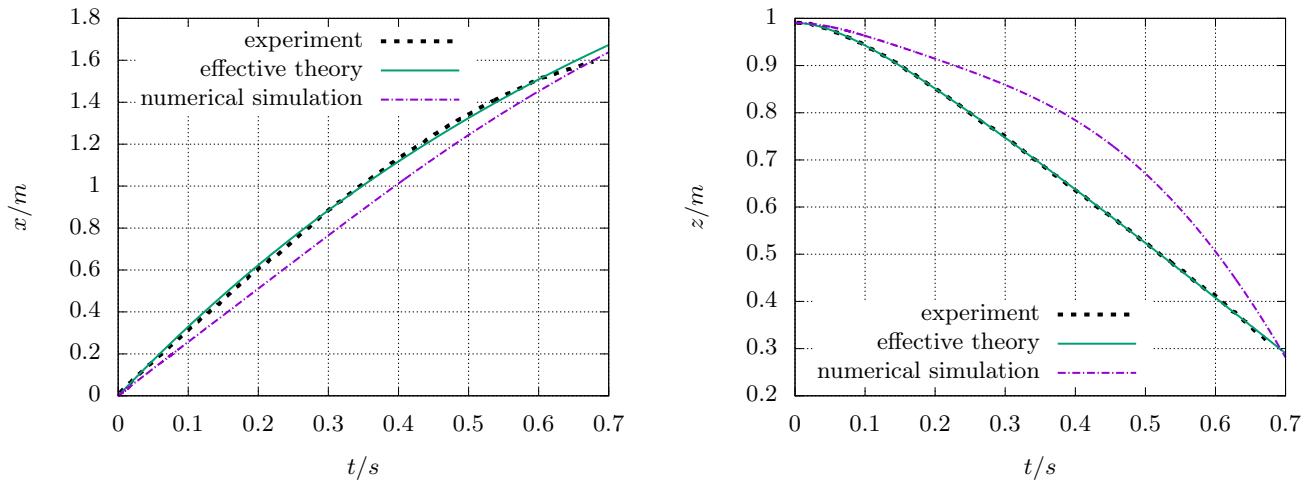
\begin{figure}[ht]
	\input{figures/x_t_5000_zu_0}
	\input{figures/z_t_5000_zu_0}
	\caption{Horizontal position in initial flight direction $x$ (left) and vertical position $z$ (right) in meters against time in seconds. The flight started with initial conditions as in eq.~\eqref{eqn_default_init_conditions} with $h=\SI{98}{\centi\meter}$, $v_0=\SI{2.6}{\meter/\second}$ and $\omega_0=\SI{49}{\radian/\second}$ (corresponding to $\num{7.9}$ rotations per second). The fit follows eq.~\eqref{eqn_eff_sol_x1} and eq.~\eqref{eqn_eff_sol_x3} respectively. For details on the numerical simulation see section~\ref{sec:numerical_simulation}. The lengths are not exactly correct due to perspective distortion.}\label{fig_x_t_trajectory}
\end{figure}

	In figure~\ref{fig_x_t_trajectory} we visualise both $x$- and $z$-components of the same real flight presented in figure~\ref{fig_y_t_trajectory} of the main article. We find that as before the experimental data is very well matched by the effective theory. The results obtained via first principle numerical simulations (see sec.~\ref{sec:numerical_simulation}) provide a decent approximation of the real course of flight by following the correct trend.

%% file: figures/x_t_5000_zu_0.tex
\begingroup
  \inputencoding{latin1}%
  \makeatletter
  \providecommand\color[2][]{%
    \GenericError{(gnuplot) \space\space\space\@spaces}{%
      Package color not loaded in conjunction with
      terminal option `colourtext'%
    }{See the gnuplot documentation for explanation.%
    }{Either use 'blacktext' in gnuplot or load the package
      color.sty in LaTeX.}%
    \renewcommand\color[2][]{}%
  }%
  \providecommand\includegraphics[2][]{%
    \GenericError{(gnuplot) \space\space\space\@spaces}{%
      Package graphicx or graphics not loaded%
    }{See the gnuplot documentation for explanation.%
    }{The gnuplot epslatex terminal needs graphicx.sty or graphics.sty.}%
    \renewcommand\includegraphics[2][]{}%
  }%
  \providecommand\rotatebox[2]{#2}%
  \@ifundefined{ifGPcolor}{%
    \newif\ifGPcolor
    \GPcolortrue
  }{}%
  \@ifundefined{ifGPblacktext}{%
    \newif\ifGPblacktext
    \GPblacktexttrue
  }{}%
  \let\gplgaddtomacro\g@addto@macro
  \gdef\gplbacktext{}%
  \gdef\gplfronttext{}%
  \makeatother
  \ifGPblacktext
    \def\colorrgb#1{}%
    \def\colorgray#1{}%
  \else
    \ifGPcolor
      \def\colorrgb#1{\color[rgb]{#1}}%
      \def\colorgray#1{\color[gray]{#1}}%
      \expandafter\def\csname LTw\endcsname{\color{white}}%
      \expandafter\def\csname LTb\endcsname{\color{black}}%
      \expandafter\def\csname LTa\endcsname{\color{black}}%
      \expandafter\def\csname LT0\endcsname{\color[rgb]{1,0,0}}%
      \expandafter\def\csname LT1\endcsname{\color[rgb]{0,1,0}}%
      \expandafter\def\csname LT2\endcsname{\color[rgb]{0,0,1}}%
      \expandafter\def\csname LT3\endcsname{\color[rgb]{1,0,1}}%
      \expandafter\def\csname LT4\endcsname{\color[rgb]{0,1,1}}%
      \expandafter\def\csname LT5\endcsname{\color[rgb]{1,1,0}}%
      \expandafter\def\csname LT6\endcsname{\color[rgb]{0,0,0}}%
      \expandafter\def\csname LT7\endcsname{\color[rgb]{1,0.3,0}}%
      \expandafter\def\csname LT8\endcsname{\color[rgb]{0.5,0.5,0.5}}%
    \else
      \def\colorrgb#1{\color{black}}%
      \def\colorgray#1{\color[gray]{#1}}%
      \expandafter\def\csname LTw\endcsname{\color{white}}%
      \expandafter\def\csname LTb\endcsname{\color{black}}%
      \expandafter\def\csname LTa\endcsname{\color{black}}%
      \expandafter\def\csname LT0\endcsname{\color{black}}%
      \expandafter\def\csname LT1\endcsname{\color{black}}%
      \expandafter\def\csname LT2\endcsname{\color{black}}%
      \expandafter\def\csname LT3\endcsname{\color{black}}%
      \expandafter\def\csname LT4\endcsname{\color{black}}%
      \expandafter\def\csname LT5\endcsname{\color{black}}%
      \expandafter\def\csname LT6\endcsname{\color{black}}%
      \expandafter\def\csname LT7\endcsname{\color{black}}%
      \expandafter\def\csname LT8\endcsname{\color{black}}%
    \fi
  \fi
    \setlength{\unitlength}{0.0500bp}%
    \ifx\gptboxheight\undefined%
      \newlength{\gptboxheight}%
      \newlength{\gptboxwidth}%
      \newsavebox{\gptboxtext}%
    \fi%
    \setlength{\fboxrule}{0.5pt}%
    \setlength{\fboxsep}{1pt}%
\begin{picture}(5040.00,3772.00)%
    \gplgaddtomacro\gplbacktext{%
      \csname LTb\endcsname
      \put(814,704){\makebox(0,0)[r]{\strut{}$0$}}%
      \csname LTb\endcsname
      \put(814,1020){\makebox(0,0)[r]{\strut{}$0.2$}}%
      \csname LTb\endcsname
      \put(814,1337){\makebox(0,0)[r]{\strut{}$0.4$}}%
      \csname LTb\endcsname
      \put(814,1653){\makebox(0,0)[r]{\strut{}$0.6$}}%
      \csname LTb\endcsname
      \put(814,1969){\makebox(0,0)[r]{\strut{}$0.8$}}%
      \csname LTb\endcsname
      \put(814,2286){\makebox(0,0)[r]{\strut{}$1$}}%
      \csname LTb\endcsname
      \put(814,2602){\makebox(0,0)[r]{\strut{}$1.2$}}%
      \csname LTb\endcsname
      \put(814,2918){\makebox(0,0)[r]{\strut{}$1.4$}}%
      \csname LTb\endcsname
      \put(814,3235){\makebox(0,0)[r]{\strut{}$1.6$}}%
      \csname LTb\endcsname
      \put(814,3551){\makebox(0,0)[r]{\strut{}$1.8$}}%
      \csname LTb\endcsname
      \put(946,484){\makebox(0,0){\strut{}$0$}}%
      \csname LTb\endcsname
      \put(1474,484){\makebox(0,0){\strut{}$0.1$}}%
      \csname LTb\endcsname
      \put(2002,484){\makebox(0,0){\strut{}$0.2$}}%
      \csname LTb\endcsname
      \put(2530,484){\makebox(0,0){\strut{}$0.3$}}%
      \csname LTb\endcsname
      \put(3059,484){\makebox(0,0){\strut{}$0.4$}}%
      \csname LTb\endcsname
      \put(3587,484){\makebox(0,0){\strut{}$0.5$}}%
      \csname LTb\endcsname
      \put(4115,484){\makebox(0,0){\strut{}$0.6$}}%
      \csname LTb\endcsname
      \put(4643,484){\makebox(0,0){\strut{}$0.7$}}%
    }%
    \gplgaddtomacro\gplfronttext{%
      \csname LTb\endcsname
      \put(209,2127){\rotatebox{-270}{\makebox(0,0){\strut{}$x/m$}}}%
      \put(2794,154){\makebox(0,0){\strut{}$t/s$}}%
      \csname LTb\endcsname
      \put(2794,3378){\makebox(0,0)[r]{\strut{}experiment}}%
      \csname LTb\endcsname
      \put(2794,3158){\makebox(0,0)[r]{\strut{}effective theory}}%
      \csname LTb\endcsname
      \put(2794,2938){\makebox(0,0)[r]{\strut{}numerical simulation}}%
    }%
    \gplbacktext
    \put(0,0){\includegraphics{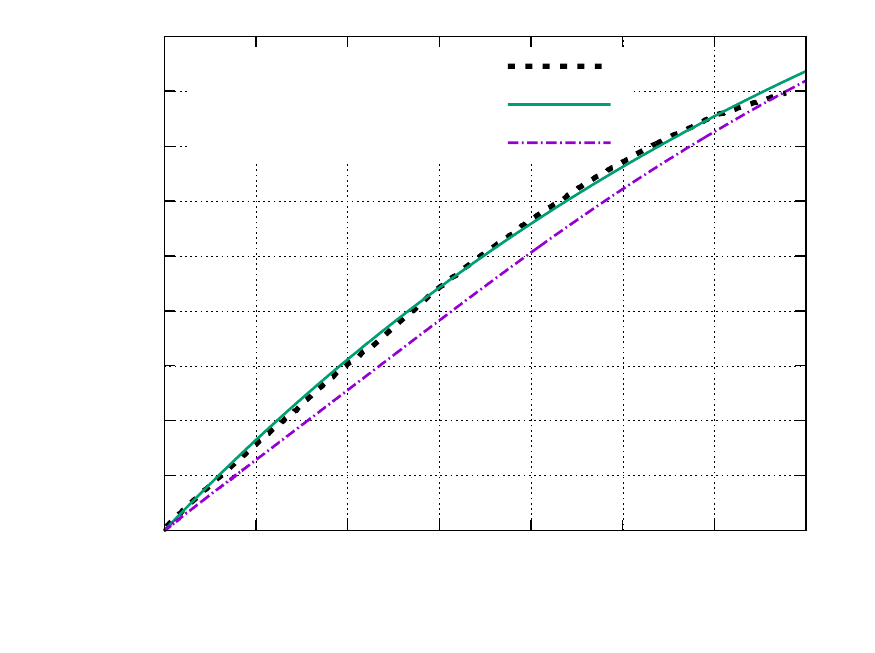}}%
    \gplfronttext
  \end{picture}%
\endgroup

%% file: figures/z_t_5000_zu_0.tex
\begingroup
  \inputencoding{latin1}%
  \makeatletter
  \providecommand\color[2][]{%
    \GenericError{(gnuplot) \space\space\space\@spaces}{%
      Package color not loaded in conjunction with
      terminal option `colourtext'%
    }{See the gnuplot documentation for explanation.%
    }{Either use 'blacktext' in gnuplot or load the package
      color.sty in LaTeX.}%
    \renewcommand\color[2][]{}%
  }%
  \providecommand\includegraphics[2][]{%
    \GenericError{(gnuplot) \space\space\space\@spaces}{%
      Package graphicx or graphics not loaded%
    }{See the gnuplot documentation for explanation.%
    }{The gnuplot epslatex terminal needs graphicx.sty or graphics.sty.}%
    \renewcommand\includegraphics[2][]{}%
  }%
  \providecommand\rotatebox[2]{#2}%
  \@ifundefined{ifGPcolor}{%
    \newif\ifGPcolor
    \GPcolortrue
  }{}%
  \@ifundefined{ifGPblacktext}{%
    \newif\ifGPblacktext
    \GPblacktexttrue
  }{}%
  \let\gplgaddtomacro\g@addto@macro
  \gdef\gplbacktext{}%
  \gdef\gplfronttext{}%
  \makeatother
  \ifGPblacktext
    \def\colorrgb#1{}%
    \def\colorgray#1{}%
  \else
    \ifGPcolor
      \def\colorrgb#1{\color[rgb]{#1}}%
      \def\colorgray#1{\color[gray]{#1}}%
      \expandafter\def\csname LTw\endcsname{\color{white}}%
      \expandafter\def\csname LTb\endcsname{\color{black}}%
      \expandafter\def\csname LTa\endcsname{\color{black}}%
      \expandafter\def\csname LT0\endcsname{\color[rgb]{1,0,0}}%
      \expandafter\def\csname LT1\endcsname{\color[rgb]{0,1,0}}%
      \expandafter\def\csname LT2\endcsname{\color[rgb]{0,0,1}}%
      \expandafter\def\csname LT3\endcsname{\color[rgb]{1,0,1}}%
      \expandafter\def\csname LT4\endcsname{\color[rgb]{0,1,1}}%
      \expandafter\def\csname LT5\endcsname{\color[rgb]{1,1,0}}%
      \expandafter\def\csname LT6\endcsname{\color[rgb]{0,0,0}}%
      \expandafter\def\csname LT7\endcsname{\color[rgb]{1,0.3,0}}%
      \expandafter\def\csname LT8\endcsname{\color[rgb]{0.5,0.5,0.5}}%
    \else
      \def\colorrgb#1{\color{black}}%
      \def\colorgray#1{\color[gray]{#1}}%
      \expandafter\def\csname LTw\endcsname{\color{white}}%
      \expandafter\def\csname LTb\endcsname{\color{black}}%
      \expandafter\def\csname LTa\endcsname{\color{black}}%
      \expandafter\def\csname LT0\endcsname{\color{black}}%
      \expandafter\def\csname LT1\endcsname{\color{black}}%
      \expandafter\def\csname LT2\endcsname{\color{black}}%
      \expandafter\def\csname LT3\endcsname{\color{black}}%
      \expandafter\def\csname LT4\endcsname{\color{black}}%
      \expandafter\def\csname LT5\endcsname{\color{black}}%
      \expandafter\def\csname LT6\endcsname{\color{black}}%
      \expandafter\def\csname LT7\endcsname{\color{black}}%
      \expandafter\def\csname LT8\endcsname{\color{black}}%
    \fi
  \fi
    \setlength{\unitlength}{0.0500bp}%
    \ifx\gptboxheight\undefined%
      \newlength{\gptboxheight}%
      \newlength{\gptboxwidth}%
      \newsavebox{\gptboxtext}%
    \fi%
    \setlength{\fboxrule}{0.5pt}%
    \setlength{\fboxsep}{1pt}%
\begin{picture}(5040.00,3772.00)%
    \gplgaddtomacro\gplbacktext{%
      \csname LTb\endcsname
      \put(814,704){\makebox(0,0)[r]{\strut{}$0.2$}}%
      \csname LTb\endcsname
      \put(814,1060){\makebox(0,0)[r]{\strut{}$0.3$}}%
      \csname LTb\endcsname
      \put(814,1416){\makebox(0,0)[r]{\strut{}$0.4$}}%
      \csname LTb\endcsname
      \put(814,1772){\makebox(0,0)[r]{\strut{}$0.5$}}%
      \csname LTb\endcsname
      \put(814,2128){\makebox(0,0)[r]{\strut{}$0.6$}}%
      \csname LTb\endcsname
      \put(814,2483){\makebox(0,0)[r]{\strut{}$0.7$}}%
      \csname LTb\endcsname
      \put(814,2839){\makebox(0,0)[r]{\strut{}$0.8$}}%
      \csname LTb\endcsname
      \put(814,3195){\makebox(0,0)[r]{\strut{}$0.9$}}%
      \csname LTb\endcsname
      \put(814,3551){\makebox(0,0)[r]{\strut{}$1$}}%
      \csname LTb\endcsname
      \put(946,484){\makebox(0,0){\strut{}$0$}}%
      \csname LTb\endcsname
      \put(1474,484){\makebox(0,0){\strut{}$0.1$}}%
      \csname LTb\endcsname
      \put(2002,484){\makebox(0,0){\strut{}$0.2$}}%
      \csname LTb\endcsname
      \put(2530,484){\makebox(0,0){\strut{}$0.3$}}%
      \csname LTb\endcsname
      \put(3059,484){\makebox(0,0){\strut{}$0.4$}}%
      \csname LTb\endcsname
      \put(3587,484){\makebox(0,0){\strut{}$0.5$}}%
      \csname LTb\endcsname
      \put(4115,484){\makebox(0,0){\strut{}$0.6$}}%
      \csname LTb\endcsname
      \put(4643,484){\makebox(0,0){\strut{}$0.7$}}%
    }%
    \gplgaddtomacro\gplfronttext{%
      \csname LTb\endcsname
      \put(209,2127){\rotatebox{-270}{\makebox(0,0){\strut{}$z/m$}}}%
      \put(2794,154){\makebox(0,0){\strut{}$t/s$}}%
      \csname LTb\endcsname
      \put(2794,1317){\makebox(0,0)[r]{\strut{}experiment}}%
      \csname LTb\endcsname
      \put(2794,1097){\makebox(0,0)[r]{\strut{}effective theory}}%
      \csname LTb\endcsname
      \put(2794,877){\makebox(0,0)[r]{\strut{}numerical simulation}}%
    }%
    \gplbacktext
    \put(0,0){\includegraphics{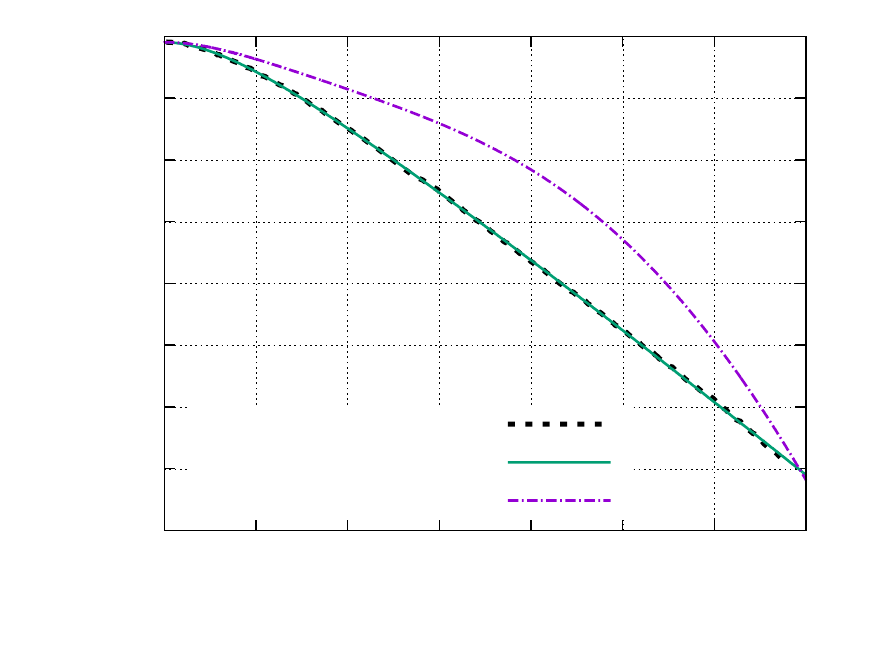}}%
    \gplfronttext
  \end{picture}%
\endgroup

%% file: sections/experimental_setup.tex
\section{Experimental setups and fit results}\label{sec_experimental_setup}

The different setups used for the experimental analysis of the beer mats' flights are listed in table~\ref{tab:sets}. The corresponding fit results can be found in table~\ref{tab:fit_results}.
The fit has been done using the Levenberg–Marquardt algorithm~\cite{levenberg,marquardt}. Ref.~\cite{lm_explained_gsl} provides a comprehensible explanation of the uncertainty estimation.

\begin{table}[h!]
	\centering
	\begin{tabular}{lS[table-format=6.1]S[table-format=-4.1]SSSS[table-format=1]}
		\hline\hline
		Set & $u_l\, \left[\text{rpm}\right]$ & $u_r\, \left[\text{rpm}\right]$ & $v_{0}\,\left[\si{\meter/\second}\right]\;$ & $\omega_{0}\,\left[\si{\radian/\second}\right]\;$ & $v_{0}/\omega_{0}\,\left[\si{\meter}\right]\;$ & $N$ \\
		\hline
		1 & 5000 & 0 & 2.6 & 49.4 & 0.05 & 1 \\ \hline
		2 & 10000 & 0 & 5.2 & 98.8 & 0.05 & 1 \\ \hline
		3 & 10000 & 5000 & 7.9 & 49.4 & 0.16 & 3 \\ \hline
		4 & 10000 & 6000 & 8.4 & 39.5 & 0.21 & 1 \\ \hline
		5 & 10500 & 5000 & 8.1 & 54.3 & 0.15 & 1 \\ \hline
		6 & 13000 & 5000 & 9.4 & 79.0 & 0.12 & 1 \\ \hline
		7 & 14000 & 6000 & 10.5 & 79.0 & 0.13 & 1 \\ \hline
		8 & 15000 & -3000 & 6.3 & 177.8 & 0.04 & 1 \\ \hline
		9 & 15000 & 0 & 7.9 & 148.2 & 0.05 & 1 \\ \hline
		10 & 15000 & 3000 & 9.4 & 118.6 & 0.08 & 1 \\ \hline
		11 & 15000 & 5000 & 10.5 & 98.8 & 0.11 & 1 \\ \hline
		\hline
	\end{tabular}
	\caption{Initial conditions of the analysed beer mat flights. $u_l$ and $u_r$ are the rotation speeds (measured in rounds per minute) of the left and right gear of the beer mat shooting apparatus respectively. The resulting velocity and angular velocity are denoted by $v_0$ and $\omega_{0}$. Each setup has been analysed $N$ times independently. We used Set 1 for all the explicit examples presented in this work.}
	\label{tab:sets}
\end{table}

\begin{table}[h!]
	\centering
	\begin{tabular}{lS[table-format=2.9]S[table-format=1.8]S[table-format=1.1]}
		\hline\hline
		Set & $\lambda\, \left[\si{\second}^{-1}\right]$ & $t_{i}\, \left[\si{\second}\right]$ & $\sqrt{\text{RSS}/\text{dof}}\, \left[\si{\milli\meter}\right]$ \\ \hline
		1  & 1.753(11) & 0.0733(12) & 5.6 \\ \hline
		2  & 1.827(19) & 0.0929(21) & 9.9 \\ \hline
		3a & 4.631(21) & 0.0377(06) & 2.5 \\ \hline
		3b & 4.140(26) & 0.0491(08) & 2.8 \\ \hline
		3c & 4.016(17) & 0.0367(07) & 2.4 \\ \hline
		4  & 7.026(101) & 0.0394(09) & 2.9 \\ \hline
		5  & 3.802(17) & 0.0420(09) & 2.6 \\ \hline
		6  & 2.902(13) & 0.0677(07) & 2.2 \\ \hline
		7  & 2.599(09) & 0.0521(08) & 1.8 \\ \hline
		8  & 1.697(09) & 0.0830(11) & 2.5 \\ \hline
		9  & 1.591(05) & 0.1062(07) & 2.2 \\ \hline
		10 & 2.045(11) & 0.0763(12) & 2.3 \\ \hline
		11 & 2.175(04) & 0.0802(05) & 1.7 \\ \hline
		\hline
	\end{tabular}
	\caption{Results of the fits following
		Eq.~\eqref{eqn_eff_sol_x2} for the initial conditions in Tab.~\ref{tab:sets}.
		Damping factor $\lambda$ and intermediate time $t_i$ are also visualised in Figs.~\ref{fig_characteristic_times} and~\ref{fig_lambdas}. The last column (root residual sum of squares over degrees of freedom) provides the standard deviation of the experimental data from the fit curve.}
	\label{tab:fit_results}
\end{table}

%% file: sections/no_spin.tex
\section{Flight with zero angular momentum}\label{app:no_spin}
	It is quite instructive to understand the unstable flight of a non-rotating disk because the influence of forces leading to precession otherwise can be observed directly. We find that the disk orientation undergoes a spontaneous symmetry breaking by tilting either slightly up- or downwards.\footnote{Example videos provided in the supplementary material.} Once it has a finite angle of attack $\alpha$, the lift-induced torque tends to increase $\alpha$. This leads to a rapid rotation and an acceleration towards the original tilting; if the movement started with an upwards tilt, the beer mat flies up, otherwise it flies down. At the same time the movement decelerates due to large drag whenever the angle of attack is different from zero.
	
\begin{figure}[ht]
		\centering
		\includegraphics[width=0.9\textwidth]{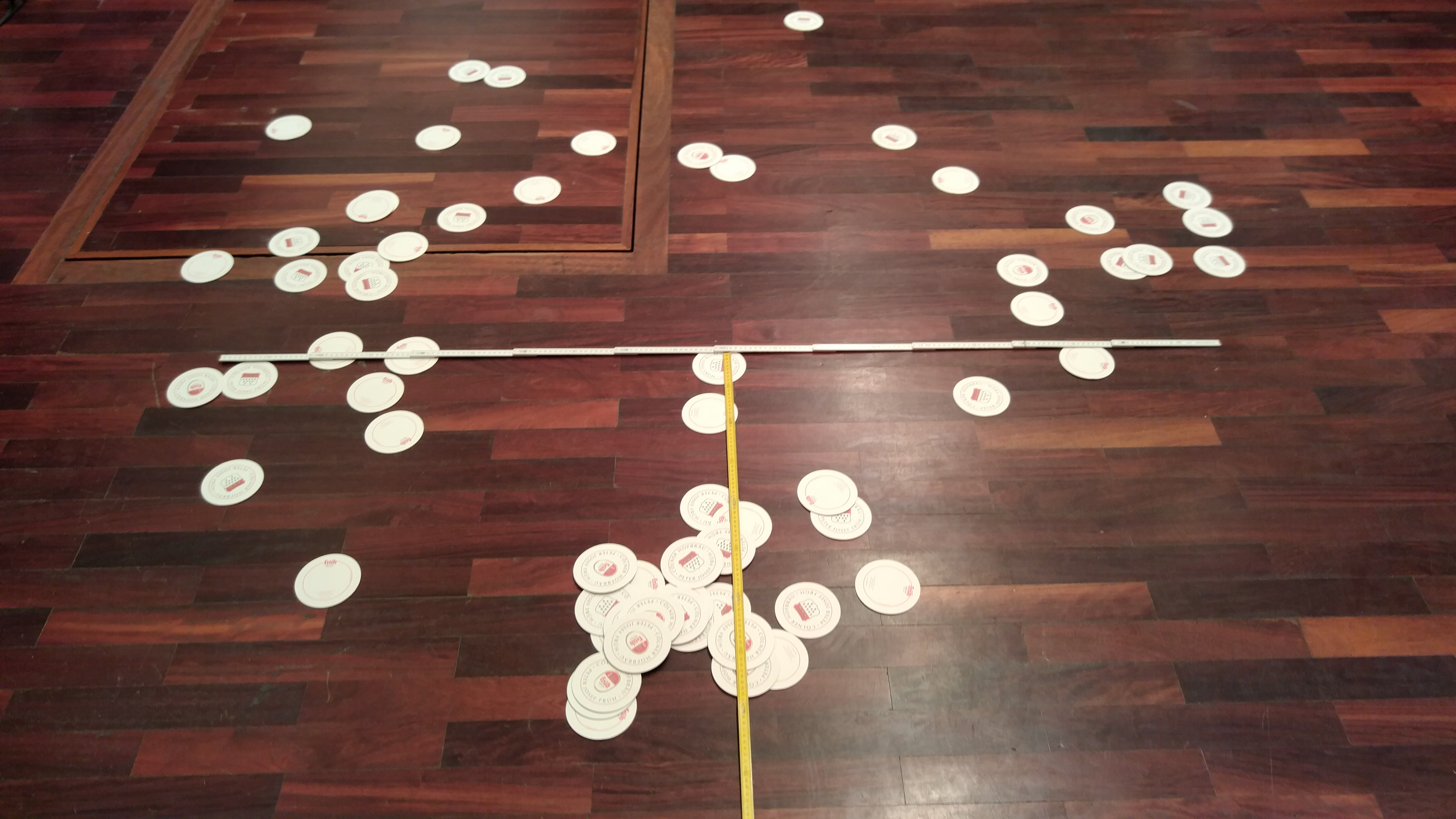}
		\caption{Scattering pattern of beer mats shot without initial angular momentum. Each folding ruler is two meters long and the flight started horizontally $\SI{98}{\centi\meter}$ above ground at $\SI{15.7}{\meter/\second}$ exactly above the unseen end of the yellow ruler.}\label{fig_no_spin}
	\end{figure}

	We now have to distinguish between the two possibilities explicitly. In the case of downwards tilt the disk flies a steep downwards curve and crashes into the floor. Successive shots hit the same place in very good approximation. There is no significant sidewards deviation from the original flight direction. This trajectory leads to the clustering in the lower central region of figure~\ref{fig_no_spin}.
	
	If, on the other hand, the disk is tilted upwards initially, it flies upwards much as a mirror image of the downwards trajectory. However, as it does not hit anything on its way, it decelerates up to some point of maximum height and nearly zero total velocity. From this point it simply falls down and minor influences can cause the disk to fall somewhere on a semicircle as can be seen in the upper half of figure~\ref{fig_no_spin}.
	
	The essential lesson from this experiment is that we can proof the existence of the lift-induced torque, or in other words show that the lift indeed attacks not in the center of mass but somewhere to the front of it.
	
	A curios reader might wonder whether upwards and downwards movements are equally likely. We repeated the experiment several times and counted the number of beer mats in the lower cluster as well as the number of beer mats in the upper semicircle. It turns out that roughly two thirds of the beer mats chose the upper trajectory and landed in the semi-circle and only one third flew downwards. A $\chi^2$-test revealed that the probability of both trajectories being equally likely lies at $p=\num{3e-8}$. Again it is clear that gravity breaks the symmetry explicitly. As the disks are accelerated downwards starting in a neutral position, it is more likely for them to end up with a positive angle of attack.

%% file: main.bbl
\begin{thebibliography}{20}%
\makeatletter
\providecommand \@ifxundefined [1]{%
 \@ifx{#1\undefined}
}%
\providecommand \@ifnum [1]{%
 \ifnum #1\expandafter \@firstoftwo
 \else \expandafter \@secondoftwo
 \fi
}%
\providecommand \@ifx [1]{%
 \ifx #1\expandafter \@firstoftwo
 \else \expandafter \@secondoftwo
 \fi
}%
\providecommand \natexlab [1]{#1}%
\providecommand \enquote  [1]{``#1''}%
\providecommand \bibnamefont  [1]{#1}%
\providecommand \bibfnamefont [1]{#1}%
\providecommand \citenamefont [1]{#1}%
\providecommand \href@noop [0]{\@secondoftwo}%
\providecommand \href [0]{\begingroup \@sanitize@url \@href}%
\providecommand \@href[1]{\@@startlink{#1}\@@href}%
\providecommand \@@href[1]{\endgroup#1\@@endlink}%
\providecommand \@sanitize@url [0]{\catcode `\\12\catcode `\$12\catcode
  `\&12\catcode `\#12\catcode `\^12\catcode `\_12\catcode `\%12\relax}%
\providecommand \@@startlink[1]{}%
\providecommand \@@endlink[0]{}%
\providecommand \url  [0]{\begingroup\@sanitize@url \@url }%
\providecommand \@url [1]{\endgroup\@href {#1}{\urlprefix }}%
\providecommand \urlprefix  [0]{URL }%
\providecommand \Eprint [0]{\href }%
\providecommand \doibase [0]{http://dx.doi.org/}%
\providecommand \selectlanguage [0]{\@gobble}%
\providecommand \bibinfo  [0]{\@secondoftwo}%
\providecommand \bibfield  [0]{\@secondoftwo}%
\providecommand \translation [1]{[#1]}%
\providecommand \BibitemOpen [0]{}%
\providecommand \bibitemStop [0]{}%
\providecommand \bibitemNoStop [0]{.\EOS\space}%
\providecommand \EOS [0]{\spacefactor3000\relax}%
\providecommand \BibitemShut  [1]{\csname bibitem#1\endcsname}%
\let\auto@bib@innerbib\@empty
\bibitem [{Cam(2013)}]{Cambridge}%
  \BibitemOpen
  \enquote {\bibinfo {title} {beer mat},}\ in\ \href
  {https://dictionary.cambridge.org/dictionary/english/beer-mat} {\emph
  {\bibinfo {booktitle} {{Cambridge Advanced Learner's Dictionary}}}}\
  (\bibinfo  {publisher} {Cambridge University Press},\ \bibinfo {year}
  {2013})\BibitemShut {NoStop}%
\bibitem [{\citenamefont {Kamaruddin}(2011)}]{frisbee_thesis}%
  \BibitemOpen
  \bibfield  {author} {\bibinfo {author} {\bibfnamefont {Noorfazreena~M.}\
  \bibnamefont {Kamaruddin}},\ }{\selectlanguage {English}\emph {\bibinfo
  {title} {{Dynamics and Performance of Flying Discs}}}},\ \href
  {https://search.proquest.com/dissertations-theses/dynamics-performance-flying-discs/docview/1775453524/se-2?accountid=14618}
  {Ph.D. thesis} (\bibinfo {year} {2011}),\ \bibinfo {note} {aerodynamic center
  of frisbees on pp. 104-112}\BibitemShut {NoStop}%
\bibitem [{\citenamefont {Hubbard}\ and\ \citenamefont
  {Hummel}(2000)}]{Hubbard00simulationof}%
  \BibitemOpen
  \bibfield  {author} {\bibinfo {author} {\bibfnamefont {M.}~\bibnamefont
  {Hubbard}}\ and\ \bibinfo {author} {\bibfnamefont {S.~A.}\ \bibnamefont
  {Hummel}},\ }\bibfield  {title} {\enquote {\bibinfo {title} {{Simulation of
  Frisbee Flight}},}\ }in\ \href
  {http://citeseerx.ist.psu.edu/viewdoc/summary?doi=10.1.1.196.2540} {\emph
  {\bibinfo {booktitle} {{5th Conference on Mathematics and Computers in
  Sport}}}},\ \bibinfo {editor} {edited by\ \bibinfo {editor} {\bibnamefont
  {{G. Cohen}}}}\ (\bibinfo {address} {Sydney, New South Wales, Australia},\
  \bibinfo {year} {2000})\BibitemShut {NoStop}%
\bibitem [{\citenamefont {Schroeder}(2015)}]{honors_theses}%
  \BibitemOpen
  \bibfield  {author} {\bibinfo {author} {\bibfnamefont {Erynn~J.}\
  \bibnamefont {Schroeder}},\ }\emph {\bibinfo {title} {{An Aerodynamic
  Simulation of Disc Flight}}},\ \href
  {https://digitalcommons.csbsju.edu/honors_theses/68} {Ph.D. thesis} (\bibinfo
  {year} {2015})\BibitemShut {NoStop}%
\bibitem [{\citenamefont {Lorenz}(2005)}]{Lorenz_2005}%
  \BibitemOpen
  \bibfield  {author} {\bibinfo {author} {\bibfnamefont {Ralph~D}\ \bibnamefont
  {Lorenz}},\ }\bibfield  {title} {\enquote {\bibinfo {title} {{Flight and
  attitude dynamics measurements of an instrumented Frisbee}},}\ }\href
  {\doibase 10.1088/0957-0233/16/3/017} {\bibfield  {journal} {\bibinfo
  {journal} {Measurement Science and Technology}\ }\textbf {\bibinfo {volume}
  {16}},\ \bibinfo {pages} {738--748} (\bibinfo {year} {2005})}\BibitemShut
  {NoStop}%
\bibitem [{\citenamefont {Motoyama}(2002)}]{Motoyama2002ThePO}%
  \BibitemOpen
  \bibfield  {author} {\bibinfo {author} {\bibfnamefont {E.}~\bibnamefont
  {Motoyama}},\ }\href
  {http://people.csail.mit.edu/jrennie/discgolf/physics.pdf} {\enquote
  {\bibinfo {title} {{The Physics of Flying Discs}},}\ } (\bibinfo {year}
  {2002}),\ \bibinfo {note} {unpublished manuscript, retrieved online June 11,
  2021}\BibitemShut {NoStop}%
\bibitem [{\citenamefont {Schlichting}\ and\ \citenamefont
  {Truckenbrodt}(1967)}]{aerodynamik}%
  \BibitemOpen
  \bibfield  {author} {\bibinfo {author} {\bibfnamefont {H.}~\bibnamefont
  {Schlichting}}\ and\ \bibinfo {author} {\bibfnamefont {E.}~\bibnamefont
  {Truckenbrodt}},\ }\href {\doibase 10.1007/978-3-642-96046-8} {\emph
  {\bibinfo {title} {Aerodynamik des Flugzeuges. Erster Band: Grundlagen aus
  der Strömungsmechanik, Aerodynamik des Tragflügels (Teil~I)}}}\ (\bibinfo
  {publisher} {Springer-Verlag},\ \bibinfo {address}
  {Berlin/Heidelberg/New~York},\ \bibinfo {year} {1967})\BibitemShut {NoStop}%
\bibitem [{\citenamefont {Abbott}\ and\ \citenamefont
  {Von~Doenhoff}(1959)}]{abbott1959theory}%
  \BibitemOpen
  \bibfield  {author} {\bibinfo {author} {\bibfnamefont {I.H.}\ \bibnamefont
  {Abbott}}\ and\ \bibinfo {author} {\bibfnamefont {A.E.}\ \bibnamefont
  {Von~Doenhoff}},\ }\href {https://books.google.de/books?id=DPZYUGNyuboC}
  {\emph {\bibinfo {title} {{Theory of Wing Sections, Including a Summary of
  Airfoil Data}}}},\ Dover Books on Aeronautical Engineering Series\ (\bibinfo
  {publisher} {Dover Publications},\ \bibinfo {year} {1959})\ \bibinfo {note}
  {aerodynamic center derived in chapter 4.}\BibitemShut {Stop}%
\bibitem [{\citenamefont {Rameka}(2009)}]{fade_explained}%
  \BibitemOpen
  \bibfield  {author} {\bibinfo {author} {\bibnamefont {Rameka}},\ }\href
  {https://www.dgcoursereview.com/forums/showthread.php?t=2250} {\enquote
  {\bibinfo {title} {{Explanation of the physics of flying discs}},}\ }
  (\bibinfo {year} {2009}),\ \bibinfo {note} {online blog, visited online June
  11, 2021}\BibitemShut {NoStop}%
\bibitem [{\citenamefont {Magnus}(1853)}]{magnus_effect}%
  \BibitemOpen
  \bibfield  {author} {\bibinfo {author} {\bibfnamefont {G.}~\bibnamefont
  {Magnus}},\ }\bibfield  {title} {\enquote {\bibinfo {title} {{Ueber die
  Abweichung der Geschosse, und: Ueber eine auffallende Erscheinung bei
  rotirenden Körpern}},}\ }\href {\doibase 10.1002/andp.18531640102}
  {\bibfield  {journal} {\bibinfo  {journal} {Annalen der Physik}\ }\textbf
  {\bibinfo {volume} {164}},\ \bibinfo {pages} {1--29} (\bibinfo {year}
  {1853})}\BibitemShut {NoStop}%
\bibitem [{\citenamefont {Muller}(2015)}]{magnus_basketball}%
  \BibitemOpen
  \bibfield  {author} {\bibinfo {author} {\bibfnamefont {Derek~Alexander}\
  \bibnamefont {Muller}},\ }\href {https://www.youtube.com/watch?v=2OSrvzNW9FE}
  {\enquote {\bibinfo {title} {{Backspin Basketball Flies Off Dam}},}\ }
  (\bibinfo {year} {2015}),\ \bibinfo {note} {youtube video, retrieved online
  June 11, 2021}\BibitemShut {NoStop}%
\bibitem [{\citenamefont {Rober}(2018)}]{playing_cards}%
  \BibitemOpen
  \bibfield  {author} {\bibinfo {author} {\bibfnamefont {Mark}\ \bibnamefont
  {Rober}},\ }\href {https://www.youtube.com/watch?v=GYCI58pMGuQ} {\enquote
  {\bibinfo {title} {{Playing Card Machine Gun- Card Throwing Trick Shots}},}\
  } (\bibinfo {year} {2018}),\ \bibinfo {note} {youtube video, retrieved online
  June 11, 2021}\BibitemShut {NoStop}%
\bibitem [{\citenamefont {Benson}(2018)}]{nasa_lift}%
  \BibitemOpen
  \bibfield  {author} {\bibinfo {author} {\bibfnamefont {Tom}\ \bibnamefont
  {Benson}},\ }\href {https://wright.nasa.gov/airplane/lifteq.html} {\enquote
  {\bibinfo {title} {{Modern Lift Equation}},}\ } (\bibinfo {year}
  {2018})\BibitemShut {NoStop}%
\bibitem [{\citenamefont {Hairer}\ and\ \citenamefont
  {Wanner}(1996)}]{wanner_ode_stiff}%
  \BibitemOpen
  \bibfield  {author} {\bibinfo {author} {\bibfnamefont {Ernst}\ \bibnamefont
  {Hairer}}\ and\ \bibinfo {author} {\bibfnamefont {Gerhard}\ \bibnamefont
  {Wanner}},\ }\href {https://archive-ouverte.unige.ch/unige:12344} {\emph
  {\bibinfo {title} {{Solving Ordinary Differential Equations II. Stiff and
  Differential-Algebraic Problems}}}}\ (\bibinfo  {publisher} {Springer},\
  \bibinfo {address} {Berlin},\ \bibinfo {year} {1996})\BibitemShut {NoStop}%
\bibitem [{\citenamefont {Dormand}\ and\ \citenamefont
  {Prince}(1980)}]{dormand_prince}%
  \BibitemOpen
  \bibfield  {author} {\bibinfo {author} {\bibfnamefont {J.R.}\ \bibnamefont
  {Dormand}}\ and\ \bibinfo {author} {\bibfnamefont {P.J.}\ \bibnamefont
  {Prince}},\ }\bibfield  {title} {\enquote {\bibinfo {title} {{A family of
  embedded Runge-Kutta formulae}},}\ }\href {\doibase
  https://doi.org/10.1016/0771-050X(80)90013-3} {\bibfield  {journal} {\bibinfo
   {journal} {Journal of Computational and Applied Mathematics}\ }\textbf
  {\bibinfo {volume} {6}},\ \bibinfo {pages} {19 -- 26} (\bibinfo {year}
  {1980})}\BibitemShut {NoStop}%
\bibitem [{\citenamefont {Hairer}\ \emph {et~al.}(1993)\citenamefont {Hairer},
  \citenamefont {N{\o}rsett},\ and\ \citenamefont
  {Wanner}}]{wanner_ode_nonstiff}%
  \BibitemOpen
  \bibfield  {author} {\bibinfo {author} {\bibfnamefont {E.}~\bibnamefont
  {Hairer}}, \bibinfo {author} {\bibfnamefont {S.~P.}\ \bibnamefont
  {N{\o}rsett}}, \ and\ \bibinfo {author} {\bibfnamefont {G.}~\bibnamefont
  {Wanner}},\ }\href@noop {} {\emph {\bibinfo {title} {{Solving Ordinary
  Differential Equations I: Nonstiff Problems}}}}\ (\bibinfo  {publisher}
  {Springer-Verlag},\ \bibinfo {address} {Berlin, Heidelberg},\ \bibinfo {year}
  {1993})\BibitemShut {NoStop}%
\bibitem [{\citenamefont {Ashino}\ \emph {et~al.}(2000)\citenamefont {Ashino},
  \citenamefont {Nagase},\ and\ \citenamefont {Vaillancourt}}]{matlab_ode}%
  \BibitemOpen
  \bibfield  {author} {\bibinfo {author} {\bibfnamefont {R.}~\bibnamefont
  {Ashino}}, \bibinfo {author} {\bibfnamefont {M.}~\bibnamefont {Nagase}}, \
  and\ \bibinfo {author} {\bibfnamefont {R.}~\bibnamefont {Vaillancourt}},\
  }\bibfield  {title} {\enquote {\bibinfo {title} {{Behind and beyond the
  Matlab ODE suite}},}\ }\href {\doibase
  https://doi.org/10.1016/S0898-1221(00)00175-9} {\bibfield  {journal}
  {\bibinfo  {journal} {{Computers \& Mathematics with Applications}}\ }\textbf
  {\bibinfo {volume} {40}},\ \bibinfo {pages} {491 -- 512} (\bibinfo {year}
  {2000})},\ \bibinfo {note} {{Our algorithm ROW23 is called ode23s
  here.}}\BibitemShut {Stop}%
\bibitem [{\citenamefont {Levenberg}(1944)}]{levenberg}%
  \BibitemOpen
  \bibfield  {author} {\bibinfo {author} {\bibfnamefont {Kenneth}\ \bibnamefont
  {Levenberg}},\ }\bibfield  {title} {\enquote {\bibinfo {title} {A method for
  the solution of certain non-linear problems in least squares},}\ }\href
  {http://www.jstor.org/stable/43633451} {\bibfield  {journal} {\bibinfo
  {journal} {Quarterly of Applied Mathematics}\ }\textbf {\bibinfo {volume}
  {2}},\ \bibinfo {pages} {164--168} (\bibinfo {year} {1944})}\BibitemShut
  {NoStop}%
\bibitem [{\citenamefont {Marquardt}(1963)}]{marquardt}%
  \BibitemOpen
  \bibfield  {author} {\bibinfo {author} {\bibfnamefont {Donald~W.}\
  \bibnamefont {Marquardt}},\ }\bibfield  {title} {\enquote {\bibinfo {title}
  {{An Algorithm for Least-Squares Estimation of Nonlinear Parameters}},}\
  }\href {\doibase 10.1137/0111030} {\bibfield  {journal} {\bibinfo  {journal}
  {Journal of the Society for Industrial and Applied Mathematics}\ }\textbf
  {\bibinfo {volume} {11}},\ \bibinfo {pages} {431--441} (\bibinfo {year}
  {1963})}\BibitemShut {NoStop}%
\bibitem [{lm_(2019)}]{lm_explained_gsl}%
  \BibitemOpen
  \bibfield  {title} {\enquote {\bibinfo {title} {{Nonlinear Least-Squares
  Fitting}},}\ }\href {https://www.gnu.org/software/gsl/doc/html/nls.html}
  {\bibfield  {journal} {\bibinfo  {journal} {GNU Scientific Library}\ }
  (\bibinfo {year} {2019})}\BibitemShut {NoStop}%
\end{thebibliography}%
